\documentclass[a4paper,fleqn]{cas-dc}

\usepackage[numbers]{natbib}
\usepackage[ruled,vlined]{algorithm2e}
\usepackage{multirow}
\usepackage{subcaption}
\usepackage{longtable}
\usepackage{supertabular}
\usepackage{tabularx}
\usepackage{multirow}
\usepackage{graphicx}
\usepackage[normalem]{ulem}
\useunder{\uline}{\ul}{}
\usepackage{float}
\def\tsc#1{\csdef{#1}{\textsc{\lowercase{#1}}\xspace}}
\tsc{WGM}
\tsc{QE}
\tsc{EP}
\tsc{PMS}
\tsc{BEC}
\tsc{DE}

\begin{document}
\let\WriteBookmarks\relax
\def\floatpagepagefraction{1}
\def\textpagefraction{.001}
\shorttitle{A Survey on the Application of Generative Adversarial Networks in Cybersecurity: Prospective, Direction and Open Research Scopes}

\title [mode = title]{A Survey on the Application of Generative Adversarial Networks in Cybersecurity: Prospective, Direction and Open Research Scopes}                      




\author[1]{Md Mashrur Arifin}
\ead{mdmashrurarifin@u.boisestate.edu}

\author[1]{Md Shoaib Ahmed}[orcid=0000-0002-6938-7309]
\cormark[1]
\ead{mdshoaibahmed@u.boisestate.edu}

\author[1]{Tanmai Kumar Ghosh}
\ead{tanmaighosh@u.boisestate.edu}

\author[1]{Ikteder Akhand Udoy}
\ead{iktederakhandudo989@u.boisestate.edu}

\author[1]{Jun Zhuang}
\ead{junzhuang@boisestate.edu}

\author[1]{Jyh-haw Yeh}
\ead{jhyeh@boisestate.edu}

\address[1]{Department of Computer Science, Boise State University, Idaho, USA}

\cortext[cor1]{Corresponding author}

\begin{abstract}
With the proliferation of Artificial Intelligence, there has been a massive increase in the amount of data required to be accumulated and disseminated digitally. As the data are available online in digital landscapes with complex and sophisticated infrastructures, it is crucial to implement various defense mechanisms based on cybersecurity. Generative Adversarial Networks (GANs), which are deep learning models, have emerged as powerful solutions for addressing the constantly changing security issues. This survey studies the significance of the deep learning model, precisely on GANs, in strengthening cybersecurity defenses. Our survey aims to explore the various works completed in GANs, such as Intrusion Detection Systems (IDS), Mobile and Network Trespass, BotNet Detection, and Malware Detection. The focus is to examine how GANs can be influential tools to strengthen cybersecurity defenses in these domains. Further, the paper discusses the challenges and constraints of using GANs in these areas and suggests future research directions. Overall, the paper highlights the potential of GANs in enhancing cybersecurity measures and addresses the need for further exploration in this field.
\end{abstract}

\begin{keywords}
Generative Adversarial Networks \sep GAN Survey \sep GAN in Cybersecurity \sep Deep Learning in Cybersecurity \sep Deep Learning Survey in Cybersecurity  \sep Survey in GAN 
\end{keywords}

\maketitle

\section{Introduction}
Cybersecurity stands as an indispensable pillar in our increasingly digital world. With the pervasive infusion of technology into every corner of our daily lives, from personal communication~\cite{mao2020perigee, xue2023goldfish} and financial transactions~\cite{baur2019cyber} to critical network infrastructure~\cite{dou2023towards, mao2023less} and national security~\cite{daricili2022national, carr2019into}, the need for robust and effective cybersecurity solutions has reached unprecedented levels. The digital landscape is fraught with an ever-evolving array of threats, including data breaches, malware, ransomware, and cyberattacks, bringing severe risks to individuals, businesses, and public interests~\cite{ali2021looking, zhuang2022geometrically, mo2022trafficflowgan}. Therefore, safeguarding our digital assets and infrastructure has become a primary concern.

Cybersecurity is confronted with many challenges, such as Adversarial Attacks~\cite{zhuang2022defending, zhuang2022robust}, Malware Detection~\cite{mazaed2022multifaceted}, Network Intrusion Detection~\cite{chen2023explainable, ruan2022causal}, etc.
First, adversarial attacks~\cite{zhou2022adversarial, lyu2024task, zhuang2022does} are a noteworthy example of these challenges, in which malevolent actors deliberately take advantage of a vulnerability in cyber-infrastructure, artificial-intelligent (AI) systems, or machine learning (ML) models. For example, adversarial attacks can inconspicuously manipulate the training or test data so that the victim models may probably misclassify instances on the test data~\cite{lyu2023backdoor, zhuang2022deperturbation}.
Second, malware~\cite{singh2021survey} is always evolving, and obfuscation techniques are used to outpace conventional detection methods, making malware detection an even more difficult task. Writers of malware are always creating new variants; one example is metamorphic malware, which changes its code structure but keeps working, avoiding detection by static analysis tools.
Third, network intrusion detection systems~\cite{liu2019machine, zhuang2021non, chen2023ride} aim to detect and prevent unauthorized access attempts, but attackers can evade the detection by using strategies like encrypted traffic and zero-day vulnerability exploitation. The increasing use of encryption in network traffic introduces a layer of complexity that makes it more difficult for intrusion detection systems~\cite{zeng2019deep, mo2022uncertainty} to detect malicious activity and analyze packet content. The difficulties highlight demands for defending the cybersecurity threats.

Generative Adversarial Networks (GANs)~\cite{goodfellow2014generative}, a type of deep-learning model~\cite{feng2018semi, feng2021two, zhang2024deepgi}, present promising solutions for tackling the above challenges~\cite{creswell2018generative}. Vanilla GANs consist of two main components: a generator and a discriminator~\cite{durgadevi2021generative, mo2022quantifying}. The generator intends to create fabricated examples that closely resemble the actual examples. On the other hand, the discriminator is specifically developed to distinguish between legitimate examples and fabricated ones.
GAN models have frequently been utilized in various cybersecurity applications over the past few years. One such application involves using GANs to strengthen machine learning models against adversarial attacks by generating adversarial samples, which can ultimately lead to improved defenses against adversarial manipulations~\cite{yan2019method}.
From the perspective of detection tasks, such as detecting malware and intrusion attempts on a network, GANs can augment threat detection mechanisms by simulating malicious behaviors, such as creating malware that can bypass antivirus or generating phishing emails that can fool both humans and machines~\cite{hu2022generating}.
In this context, integrating GAN-based models into cybersecurity strategies holds great promise, offering innovative solutions to the complex challenges that continue to evolve in the digital age.

In this work, we aspire to deliver a comprehensive exploration of the various applications and implications of GAN-based models in cybersecurity. Particularly, we investigate how GAN-based models address the following cyber threats: malware detection, anomaly detection, intrusion detection, etc. Our main objective is to provide readers with a comprehensive overview of the current development of GAN-based models in cybersecurity, including a breakdown of challenges, identification of common trends, and discussion on future directions, that serve as a guide for future research and real-world developments using GAN-based models to tackle new cybersecurity issues.

In summary, our contributions to this survey oscillate in diverse cybersecurity domains, including but not limited to malware detection, intrusion detection, anomaly detection, and detecting specific attacks like adversarial attacks, where Generative Adversarial Networks (GANs) can provide novel solutions. The main contribution can be outlined as follows:.
\begin{itemize}
    \item We systematically review the most recent literature on applying Generative Adversarial Networks (GANs) to address cyber threats.  
    \item We unveil a new taxonomy designated in this survey. Our proposed taxonomy provides readers with a clear organization to navigate the wide range of implications of GAN-based models in cybersecurity.
    \item We provide diverse figures and tables to summarize our topics. The summarization can help readers easier understand the structure of this work.
    \item We further discuss important discoveries, limitations, and new directions. The insightful discussion could inspire readers on future research directions.
\end{itemize}

\begin{figure*}
\centering
\includegraphics[scale=.65]{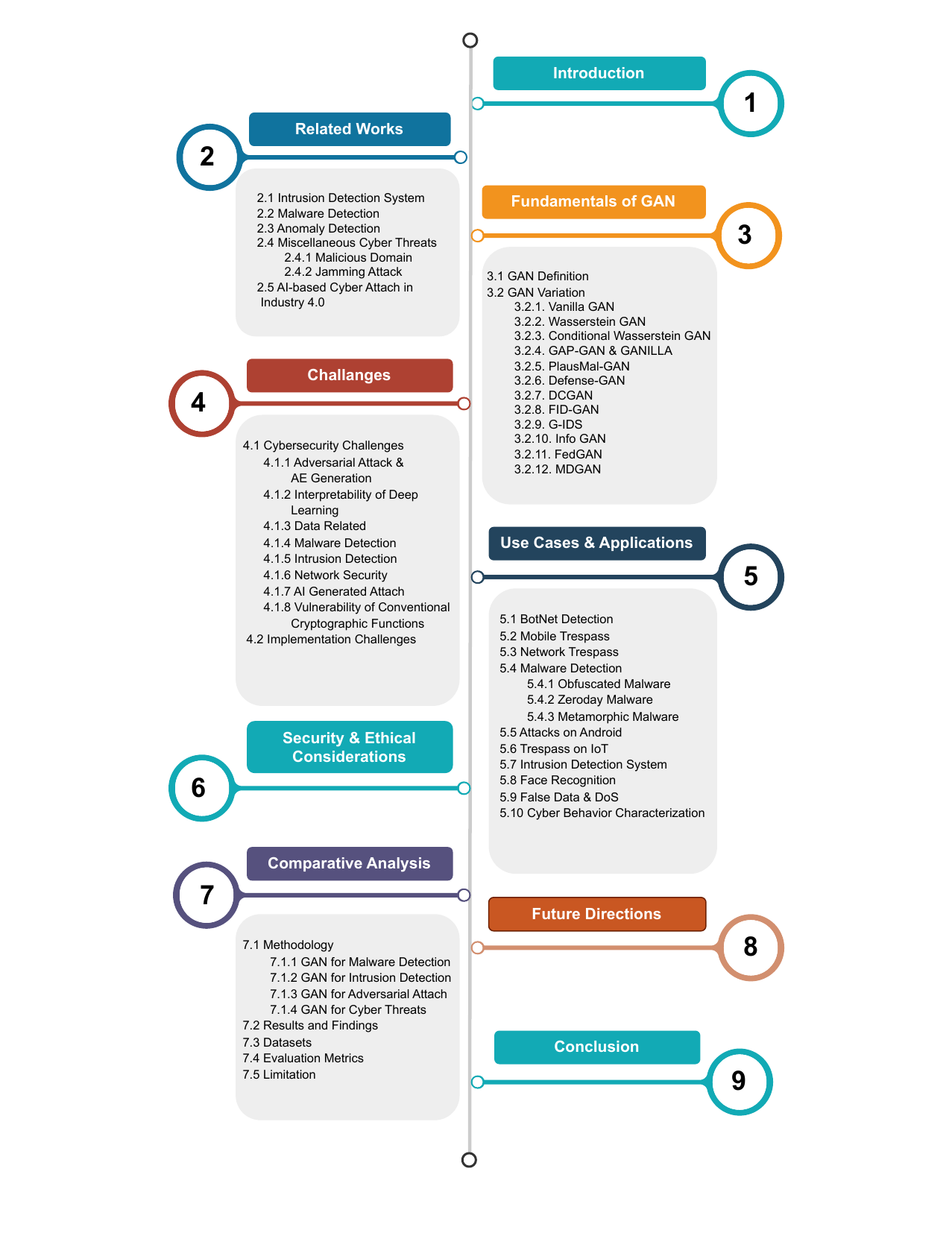}
\caption{A comprehensive overview of the state-of-the-art research on generative adversarial networks (GANs) and their applications in cybersecurity into nine sections. Each section covers a different aspect of the topic, such as the fundamentals of GANs, the cybersecurity challenges they pose and solve, the use cases and applications they enable, the security analysis they require, and the future directions they suggest. The diagram shows the survey's hierarchical structure, the sections' logical flow, and each subsection's main points.}
\label{fig:section_organization}
\end{figure*}

This survey is thoughtfully structured, with clear and concise sections and sub-sections, accompanied by clear figures and tables. Figure~\ref{fig:section_organization} provides a complete overview of the survey's structure precisely, allowing readers to navigate the paper quickly and easily. Table~\ref{table:acronyms} exemplifies the meanings of essential acronyms employed throughout this survey.

\begin{table}
\caption{A glossary of the key acronyms used throughout this paper, along with their full names and meanings. The acronyms are related to the topic of generative adversarial networks (GANs) and their applications in cybersecurity.}
\label{table:acronyms}
\resizebox{\columnwidth}{!}{%
\fontsize{14}{18}\selectfont 
\begin{tabular}{|l|l|} \hline 

\textbf{Acronyms} & \textbf{Definitions}             \\\hline \hline  
GAN               & Generative Adversarial Network \\ \hline  
IDS               & Intrusion Detection System  \\ \hline  
NIDS                & Network Intrusion Detection System\\ \hline 
PE                & Portable Executable  \\ \hline  
WGAN              & Wasserstein GAN              \\ \hline  
WCGAN             & Conditional Wasserstein GAN   \\ \hline  
CPSs            & Cyber-Physical Systems  \\ \hline  
IoT             & Internet of Things \\ \hline
FedGAN/FGAN      & Federated Generative Adversarial Network \\ \hline  
DCGAN            & Deep Convolutional Generative Adversarial Network \\ \hline  
AAM-GAN          & Adversarial Attack Method Based on GAN \\ \hline
DGAN          & Domain Generation Algorithms \\ \hline
DL          & Deep Learning \\ \hline
G          & Generator \\ \hline
D          & Discriminator \\ \hline
ML          & Machine Learning \\ \hline
AEs         & Adversarial Examples \\ \hline
DNN         & Deep Neural Network \\ \hline
FDI         & False Data Injection \\ \hline
PMUs        & Phasor Measurement Units \\ \hline
RF          & Random Forest \\ \hline
DT          & Decision Tree \\ \hline
SVM         & Support Vector Machine \\ \hline
GenAI       & Generative AI \\ \hline
DoS         & Denial-of-Service \\ \hline
VAEs        & Variational Autoencoders \\ \hline
CAM         & Class Activation Map \\ \hline
AT          & Adversarial Training \\ \hline
IoV         & Internet of Vehicles \\ \hline
QSVM        & Quantum Support Vector Machine \\ \hline
SIoT        & Social Internet of Things \\ \hline
NIDS        & Network Intrusion Detection System \\ \hline
PMUs        & Phasor Measurement Units \\ \hline
MDGAN       & Multifaceted Deep Generative Adversarial Networks Mode \\ \hline
NID         & Network Intrusion Detection \\ \hline
MITM        & Man-in-the-Middle \\ \hline
BFAM        & Brute-Force Attack Method \\ \hline
ADFL        & Adversarial Distillation-based Backdoor Defense Scheme for Federated Learning \\ \hline
SMOTE       & Synthetic Minority Oversampling Technique \\ \hline
CTI         & Cyber Threat Intelligence \\ \hline
BTD         & Binary-to-DNN \\ \hline
TTPs        & Tactics, Techniques, and Procedures \\ \hline
SDN         & Software-Defined Networking \\ \hline
\end{tabular}%
}
\end{table}

\section{Related Works}
The section is responsible for presenting a thorough examination of the surveys conducted on GANs in the cybersecurity domain, with a specific focus on their use in areas like Intrusion Detection Systems, Malware Detection, and Miscellaneous Cyber Threats. We will give a brief outline of the research papers relevant to the topic and the domains they cover. We will also compare our study with other papers in the same field. The aim is to present a thorough and informative analysis of the current state of research on GANs in cybersecurity to identify trends, challenges, and opportunities for further research.

\subsection{Intrusion Detection System}
Cybersecurity is crucial to protect the increasing amount of stored data and its transmission on networks. Regular updates to threat detection techniques are necessary to prevent advanced attacks~\cite{t15_arora2022review, chen2023real}. Using GANs for Intrusion Detection Systems (IDS) introduces a novel and promising approach to enhancing cybersecurity. The survey~\cite{AS2_dunmore2023comprehensive} aims to investigate the potential use of GANs in cybersecurity, specifically for intrusion detection systems (IDS). The researchers conducted a comprehensive survey of current research and methods related to GANs in IDS, including topics such as model privacy, attack evasion, dataset evaluation, and various GAN model variants. However, this research focuses primarily on the use of GANs in IDS and does not extensively cover other areas of cybersecurity.

\subsection{Malware Detection}
Conventional malware detections often rely on signature-based methods or behavioral analysis, which can be limited in their ability to identify polymorphic or previously unseen malware variants~\cite{U9_moti2021generative, li2024comprehensive}. GAN is promising for Malware Detection since it can identify previously unseen malware variants through its ability to analyze patterns and generate new ones. 
Ling et al.~\cite{U1_ling2023adversarial} provide a detailed overview of the latest adversarial attacks on Windows PE malware detection and the corresponding defenses. The authors discuss three significant challenges in maintaining the meaning of adversarial PE malware to create practical and realistic adversarial attacks, namely format-preservation, executability-preservation, and maliciousness-preservation. Recent research efforts in adversarial attacks and defenses have been reviewed, and an organized taxonomy has been developed to summarize the existing literature. The authors suggest that researchers should focus on developing more effective and efficient defenses against adversarial attacks, exploring new techniques for generating and detecting adversarial examples, and investigating the use of reinforcement learning and other advanced machine learning techniques to improve the robustness of malware detection systems.

\subsection{Anomaly Detection}
Unsupervised learning~\cite{ma2023learning} or clustering techniques~\cite{ma2022learning} are widely used for anomaly detection, whereas GANs are one of the cutting-edge approaches within these categories, offering a range of advantages in identifying unusual patterns or behaviors in diverse datasets. For example, Xia et al.~\cite{U4_xia2022gan} propose a GAN-based unsupervised learning algorithm for detecting anomalies. This study indicates that GANs can identify abnormal patterns by learning to represent samples through adversarial learning. The paper provides three criteria for discussing anomaly detection and covers the theoretical and technological advancements, theoretical foundation, applicable tasks, and practical applications of anomaly detection based on GAN. Additionally, the paper delves into the present obstacles and difficulties faced by anomaly detection based on GANs and proposes probable paths for future research. The study compiles information from more than 330 sources regarding anomaly detection using GANs and offers in-depth technical details for scholars wishing to utilize GANs for such tasks.

\subsection{Miscellaneous Cyber Threats}
\subsubsection{Malicious Domain Detection}
The research paper~\cite{AS5_berman2019survey} highlights the importance of Generative Adversarial Networks (GANs) in the field of cybersecurity. GANs are used to overcome the challenge of identifying malicious domain names generated through Domain Generation Algorithms (DGAs). They prove to be a valuable tool in enhancing the detection of malicious domain names generated by DGAs. GANs produce synthetic data that leads to more effective model training and improved accuracy in identifying malicious domains~\cite{wang2024balanced}. 

However, the survey has some limitations. It identifies the need for developing new datasets and approaches for Deep Learning (DL) in cybersecurity but does not provide specific recommendations or guidelines on how to address these needs. 

The survey indicates that there is a need for further research in the development of new datasets. These datasets should take into account the perspective of potential adversaries and consider how they could use DL techniques to evade detection mechanisms. Comparing the performance metrics of DL methods in real operational cases is crucial, according to the survey. It emphasizes the significance of testing DL techniques in real-life situations to assess their efficiency. Furthermore, the survey highlights the importance of having extensive benchmark datasets that are frequently updated to promote the progress of cybersecurity solutions and establish a strong sense of confidence in DL techniques.

\subsubsection{Jamming Attack Detection}
Wu et al. propose two GAN-based methods to identify network attacks~\cite{AS6_wu2020network}. The first method aims to detect jamming attacks in wireless communication by using a classifier to predict channel states and determine whether to transmit based on sensing outcomes. The second approach focuses on identifying URL-based phishing attacks by leveraging generative adversarial deep neural networks. 

\subsection{AI-based Cyber Attack in Industry 4.0}
Besides reviewing GAN-related literature on AI-based cyber-attacks and their crunch on Industry 4.0, this paper~\cite{de2023artificial} discussed categories of cyber-attacks, defense countermeasures, ML and DL techniques, and the pros and cons of AI from cybersecurity perspectives in Industry 4.0. The paper also provides insights for the researchers to develop effective defenses against AI-based cyber threats. It emphasized the need to constantly update cyber security actions in response to the evolving cybernetic technological landscape. 

Our comparative analysis juxtaposes our study against the previously discussed research endeavors, each focusing on specific facets of cybersecurity problems. Notably, while these earlier studies delved into individual areas within the cybersecurity domain, our research takes a holistic approach by comprehensively addressing all these facets. The ensuing Table~\ref{tab:comparison_related_works} provides an illuminating comparison, highlighting the breadth and inclusivity of our study in contrast to these state-of-the-art surveys, thus underscoring the comprehensive nature of our contributions to the field.

\begin{table*}[t]
    \centering
    \caption{The table depicts a comprehensive comparison between our study and state-of-the-art surveys.}
    \begin{tabular}{|p{2.5cm}|p{2cm}|p{2cm}|p{2cm}|p{2cm}|p{2cm}|p{2cm}|}
    \hline
    \textbf{Authors} & \textbf{Intrusion Detection System} & \textbf{Malware Detection} & \textbf{Anomaly Detection} & \textbf{Malicious Domain Detection} & \textbf{Jamming Attack Detection} & \textbf{AI-based Cyber Attack in Industry 4.0}  \\
    \hline
    Dunmore et al.\cite{AS2_dunmore2023comprehensive} & \checkmark &  &  &  &  &   \\
    \hline
    Ling et al.~\cite{U1_ling2023adversarial} &  & \checkmark &  &  &  &   \\
    \hline
    Xia et al.\cite{U4_xia2022gan} &  &  & \checkmark &  &  &   \\
    \hline
    Berman et al.\cite{AS5_berman2019survey} &  &  &  & \checkmark &  &   \\
    \hline
    Wu et al.\cite{AS6_wu2020network} &  &  &  &  & \checkmark &   \\
    \hline
    De et al.\cite{de2023artificial} &  &  &  &  &  & \checkmark  \\
    \hline
    Our Study & \checkmark & \checkmark & \checkmark & \checkmark & \checkmark & \checkmark  \\
    \hline
    \end{tabular}
    \label{tab:comparison_related_works}
\end{table*}

\section{Fundamentals of Generative Adversarial Networks (GANs)}
In the ever-evolving realm of cybersecurity, employing cutting-edge technology is essential to maintaining an advantage over adversaries. In 2014, Goodfellow et al. first introduced vanilla GANs~\cite{goodfellow2014generative}, which has garnered substantial attention in recent years. 

\subsection{GANs Defination}
\begin{figure}
\centering
 \includegraphics[width=1\linewidth]{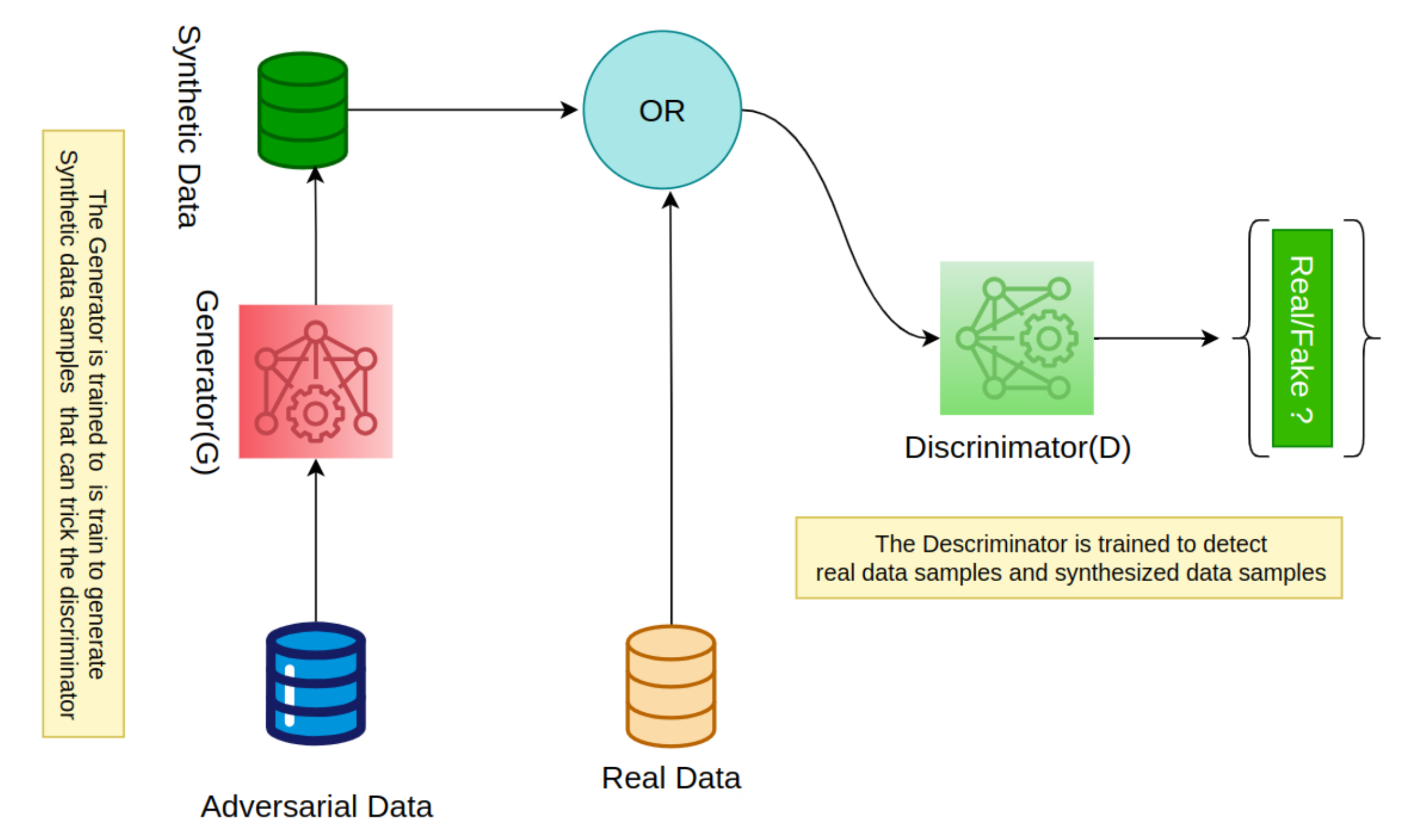}
\caption{This diagram illustrates the fundamental design of a Generative Adversarial Network (GAN). The two primary components of the GAN structure are the Generator and the Discriminator. The Generator creates fictitious data, and the Discriminator verifies the authenticity of the samples it creates. The Generator and Discriminator keep getting better at what they do through a process of constant adversarial training. This makes the data they produce more realistic.}
\label{fig:GAN_architecture}
\end{figure}

GANs are widely used in generative tasks. The generation capacity can be achieved by leveraging the strong representation of deep neural networks. As depicted in Figure~\ref{fig:GAN_architecture}, vanilla GANs consist of two neural networks, a generator ($G$) and a discriminator ($D$). During the training process, the generator $G$ takes random noise as input and generates fake samples. In the meanwhile, the discriminator $D$ distinguishes the difference between generated and ground-truth samples. Both neural networks are playing a zero-sum game during this process. After the loss curve is converged, the model reaches the Nash Equilibrium. The total loss function can be formulated as follows:
\[
\mathcal{L}_{\text{GAN}} = \mathbb{E}_{\mathbf{x} \sim p_{\text{data}}(\mathbf{x})}[\log D(\mathbf{x})] + \mathbb{E}_{\mathbf{z} \sim p_{\mathbf{z}}(\mathbf{z})}[\log(1 - D(G(\mathbf{z})))],
\]
where $\mathbf{x}$ and $\mathbf{z}$ denote real data and random noise, respectively. $p_{\text{data}}(\mathbf{x})$ and $p_{\mathbf{z}}(\mathbf{z})$ denote the distribution of real data and random noise, respectively. Rectified linear units (ReLU) are commonly used as an activation function to capture the non-linear relationship. Overall, GANs have revolutionized the field of generative modeling by introducing a novel adversarial training paradigm.

\subsection{GAN Variation}
\paragraph{\textbf{Vanilla and Conditional GAN}}
Due to the extraordinary performance of GAN, variants of GAN have been widely studied~\cite{balaji2019conditional}. Conditional GAN~\cite{balaji2019conditional} is one of such variants that incorporates supplementary information, such as labels, into generating data.

Randhawa et al.~\cite{AS1_randhawa2021security} employ two different GAN architectures for generating botnet traffic. The first architecture is the \textbf{vanilla GAN}, where the generator model G generates samples from a noise space, and the discriminator model D evaluates the generated samples. The estimation of probability for generated and real data forms the basis of the loss function for the combined model. The provided document does not go into great detail about the second architecture, which is the conditional GAN. The study notes that further research needs to be conducted to determine an appropriate GAN for producing botnet data. 

\paragraph{\textbf{Wasserstein GAN (WGAN)}}
Duy et al.~\cite{t1_duy2021digfupas} propose a variant of GANs, namely Wasserstein GAN (WGAN), that derives from using Wasserstein distance during optimization, to enhance the training stability of GAN-based models. The goal of WGAN is to use an input dataset to train a generative model that can efficiently generate adversarial samples. In the WGAN training process, the discriminator and generator compete in a minimax objective in which the discriminator attempts to identify fake samples while the generator uses its generated samples to deceive it. Wasserstein GAN (WGAN)~\cite{AS3_zhang2020brute} has shown excellent performance among various GAN variants. The research~\cite{AS7_kim2022obfuscated} incorporates Wasserstein GAN, which measures distribution differences as Wasserstein distance, in contrast to the conventional GAN that employs Jensen-Shannon distance. By creating virtual malware for detector training, this GAN-based method overcomes the problem of malware obfuscation and initializes the generative model. The primary objective is to produce modified data that exhibits resilience against malware deformations, thereby extending the detector's capability to identify a broader spectrum of malware. This innovative approach not only enables the encoder and detector to withstand modifications but also facilitates training with a more diverse dataset, enhancing the robustness and adaptability of the detector. 

\begin{figure}
\centering
\includegraphics[width=1\linewidth]{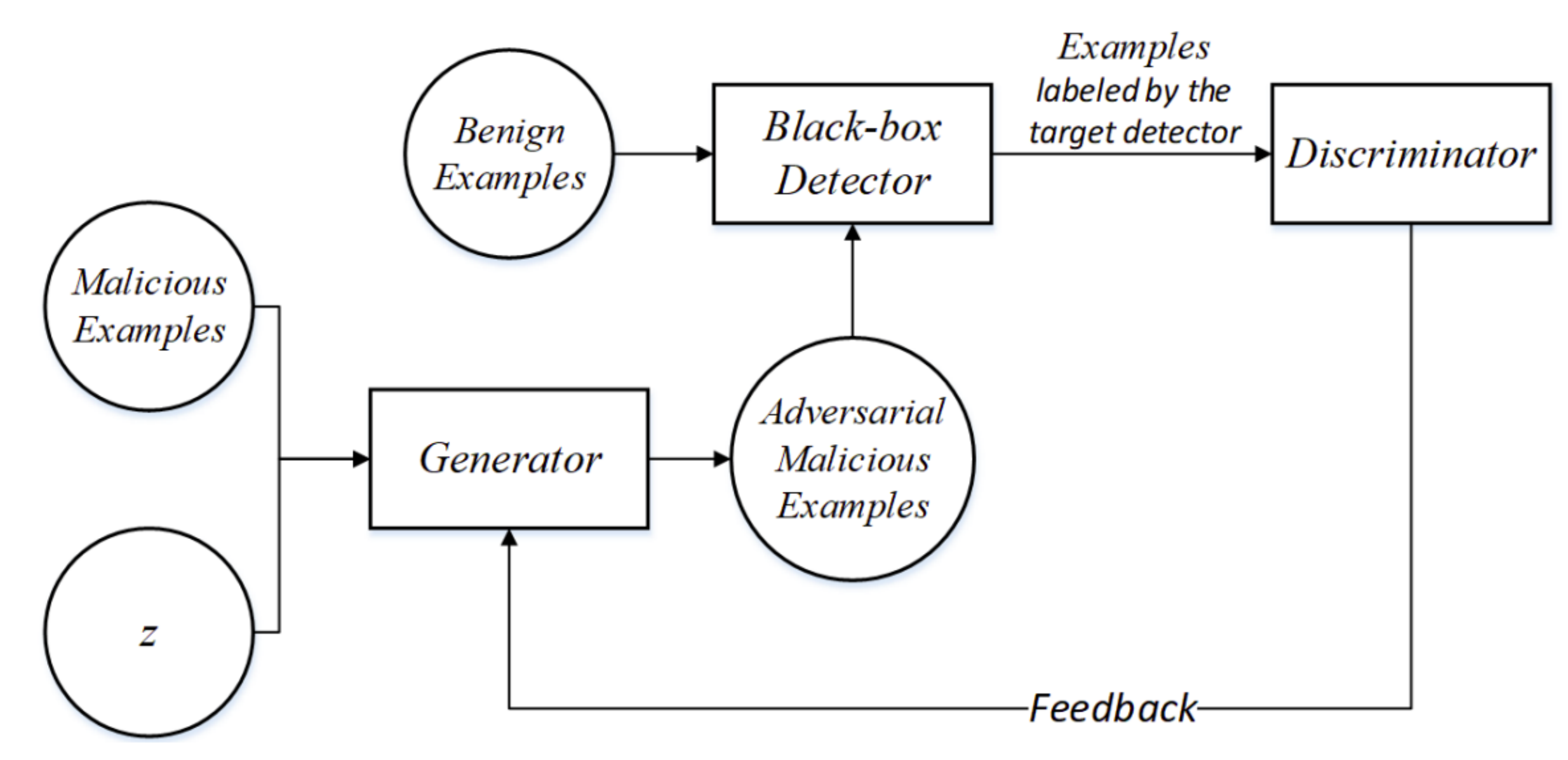}
\caption{Adversarial attack method based on GAN (AAM-GAN)~\cite{AS3_zhang2020brute}}
\label{fig:AAM-GAN}
\end{figure}

\paragraph{\textbf{Conditional Wasserstein GAN}}
The conditional Wasserstein Generative Adversarial Network (GAN) has been used in this study~\cite{AS4_chui2023three} to generate additional training data for minority classes in the network intrusion detection (NID) model. The GAN is improved to enhance its performance in generating realistic and diverse synthetic data. The generated data is then used to balance the imbalanced ratios between different types of network intrusions. The conditional Wasserstein GAN is combined with a cost-sensitive stacked autoencoder to extract features and construct the NID model. Benchmark datasets are used to assess the model's performance, and the accuracy obtained is reported. The model~\cite{AS11_kumar2023synthetic} combines an XGBoost Classifier and a Wasserstein Conditional GAN (WCGAN). To address the problem of data imbalance, the WCGAN is used to generate synthetic data samples for minority attack classes. The discriminator serves as a mentor for the generator, which is trained to generate realistic data samples that match the real training attack data samples. Promising results are observed when the performance of the suggested model has been evaluated and compared with other GAN models.

\paragraph{\textbf{GAPGAN and GANILLA}}
The application of GANs to adversarial attacks against Windows PE malware detection is covered in the paper~\cite{U1_ling2023adversarial}. The paper specifically mentions GAPGAN and GANILLA, two GAN-based attack frameworks. Black-box adversarial attack framework GAPGAN creates byte-level adversarial payloads to counter MalConv and other DL-based malware detection systems. The discriminator tries to mimic the black-box PE malware detector in order to detect both the generated adversarial PE malware and the original PE goodware. The generator creates adversarial payloads that are appended to original malware samples during the training process. Conversely, GANILLA is a white-box adversarial attack framework that modifies the original PE malware samples to produce adversarial ones. The generator creates adversarial PE malware samples by adding to or altering the original PE malware samples, and the discriminator tries to discern between the adversarial and original PE malware samples. The generator and discriminator are trained simultaneously. Compared to current techniques, the authors assert that GANILLA can produce adversarial PE malware samples that are more realistic.

\paragraph{\textbf{PlausMal-GAN}}
 Won et al.~\cite{AS8_won2022plausmal} introduce a GAN-based framework, PlausMal-GAN, that can detect zero-day malware by generating high-quality and diverse analogous malware images. PlausMal-GAN addresses the challenges of data insufficiency and imbalance in zero-day malware detection, producing reliable accuracy performances and outperforming previous methods. Several GAN models are integrated into the framework, such as DCGAN, LSGAN, WGAN-GP, and E-GAN. Each model is assessed by a discriminator and the generator that performs the best moves on to the next phase. PlausMal-GAN's methodology involves training a generator and discriminator with malware data, encompassing two key steps: initial training using real and generated malware data and re-training with real and generated data. This framework demonstrates its effectiveness in detecting and predicting numerous analogous zero-day malware instances, showcasing superior accuracy and providing consistent results across malware classification metrics. In summary, PlausMal-GAN serves as a robust malware training framework underpinned by GAN, elevating zero-day malware detection with its ability to generate high-quality, diverse malware samples and deliver dependable performance, even when confronted with limited training data, surpassing prior approaches in terms of accuracy and classification results.

\paragraph{\textbf{Defense-GAN}}
Reilly et al.~\cite{AS9_reilly2023robustness} present Defense-GAN in the context of protecting classifiers against adversarial attacks in malware classification. Defense-GAN, introduced by Samangouei et al., is a framework that utilizes generative models to enhance the robustness of classifiers. Defense-GAN has demonstrated efficacy against black-box and white-box attacks on the MNIST and F-MNIST datasets, according to the paper. By using GAN-generated samples for training, the researchers hope to strengthen a model's resilience in the context of malware classification.

\paragraph{\textbf{Deep Convolutional Generative Adversarial Network (DCGAN)}}
Lu et al.~\cite{U8_lu2019generative} state challenges of deep learning-based malware classification tasks. A deep model needs a large amount of training data to be trained. To augment the training data and raise the classifier's classification accuracy, the authors suggest a method that creates synthetic malware samples using a Deep Convolutional Generative Adversarial Network (DCGAN). The effectiveness of the suggested method in enhancing classification accuracy is demonstrated by evaluating it on the Malimg dataset and comparing it with other cutting-edge techniques.

\paragraph{\textbf{FID-GAN}}
The need for efficient and effective intrusion detection systems (IDS) for Cyber-Physical Systems (CPSs) that can meet low-latency requirements and detect new cyber-attacks without labeled data was addressed in the paper~\cite{de2020intrusion}. FID-GAN, a unique fog-based, unsupervised IDS for CPSs utilizing Generative Adversarial Networks (GANs), is the suggested remedy. In order to meet low-latency requirements, FID-GAN computes a reconstruction loss based on the reconstruction of data samples mapped to the latent space. Two datasets are used to evaluate the proposed architecture, and the findings demonstrate that FID-GAN performs better than current state-of-the-art techniques in terms of false alarm rate and detection accuracy. Non-labeled data can also be used by FID-GAN to identify new cyberattacks. FID-GAN is an unsupervised approach to detecting cyberattacks by implicitly modeling the system using generative adversarial networks (GANs). Unlike supervised approaches that require labeled data for training, unsupervised approaches can detect novel attacks without labeled data.

\paragraph{\textbf{G-IDS}}
In this study~\cite{t6_shahriar2020g}, the GAN variant used is called G-IDS, which stands for Generative Intrusion Detection System. G-IDS is a framework that utilizes a generative adversarial network (GAN) to generate more training data for an imbalanced dataset in cyber-physical systems (CPS) security. The G-IDS framework improves the performance of the intrusion detection system (IDS) by addressing the limitations of insufficient and imbalanced data.

\paragraph{\textbf{Info GAN}}
Chen et al.~\cite{chen2016infogan} present a variant of GAN, Info-GAN, that facilitates interpretable representation learning by optimizing the information present in the adversarial nets that are produced. With the aid of this GAN variation, image content can be represented in a high-dimensional manner and then reduced to new images, enabling the classifier to distinguish between important pieces of evidence that are similar and different.

\paragraph{\textbf{Federated Generative Adversarial Network (FedGAN)}}
The paper~\cite{t9_ferrag2021federated} mentions the use of a variant of Generative Adversarial Network (GAN) called Federated Generative Adversarial Network (FedGAN). FedGAN is utilized in the Internet of Things (IoT) context of distributed intrusion detection. An particular kind of GAN is used to produce data from adversarial attacks and categorize it. The discriminator and the generator are the two parts of the FedGAN. A particular type of Generative Adversarial Networks (GANs) is used in the work of Das et al.~\cite{t11_das2022fgan} to detect anomalies in network traffic. It tackles the difficulties of teaching GANs to identify anomalies in big networks with a variety of traffic types as well as the requirement for dataset privacy. Because FGAN uses a federated architecture, various network users can train and modify an adversarial model that is centrally available based on their unique circumstances. This approach enables localized and coordinated training, enhancing anomaly detection in medium to large-sized networks.

\paragraph{\textbf{MDGAN}}
MDGAN, which stands for "Multifaceted Deep Generative Adversarial Networks Model,"~\cite{mazaed2022multifaceted} is a new way to improve mobile security by tackling the danger of malware attacks. MDGAN is designed to find malware on mobile devices. It uses a multifaceted input method that blends 2D grayscale image features with LSTM (Long Short-Term Memory) binary sequence features. This one-of-a-kind approach makes it possible to find different kinds of malware, which makes the model more flexible. The skeleton is organized into three main parts. First, the raw APK package is preprocessed to create an API sequence file and a binary image. Subsequently, features are extracted from the binary image using GoogleNet, and stable features are obtained from the API sequence file using an LSTM network. Then, these features are joined together to make a single multi-face input for the model. Notably, MDGAN uses a GAN as part of its design to improve the training data. 

\begin{figure*}
\centering
\includegraphics[scale=.32]{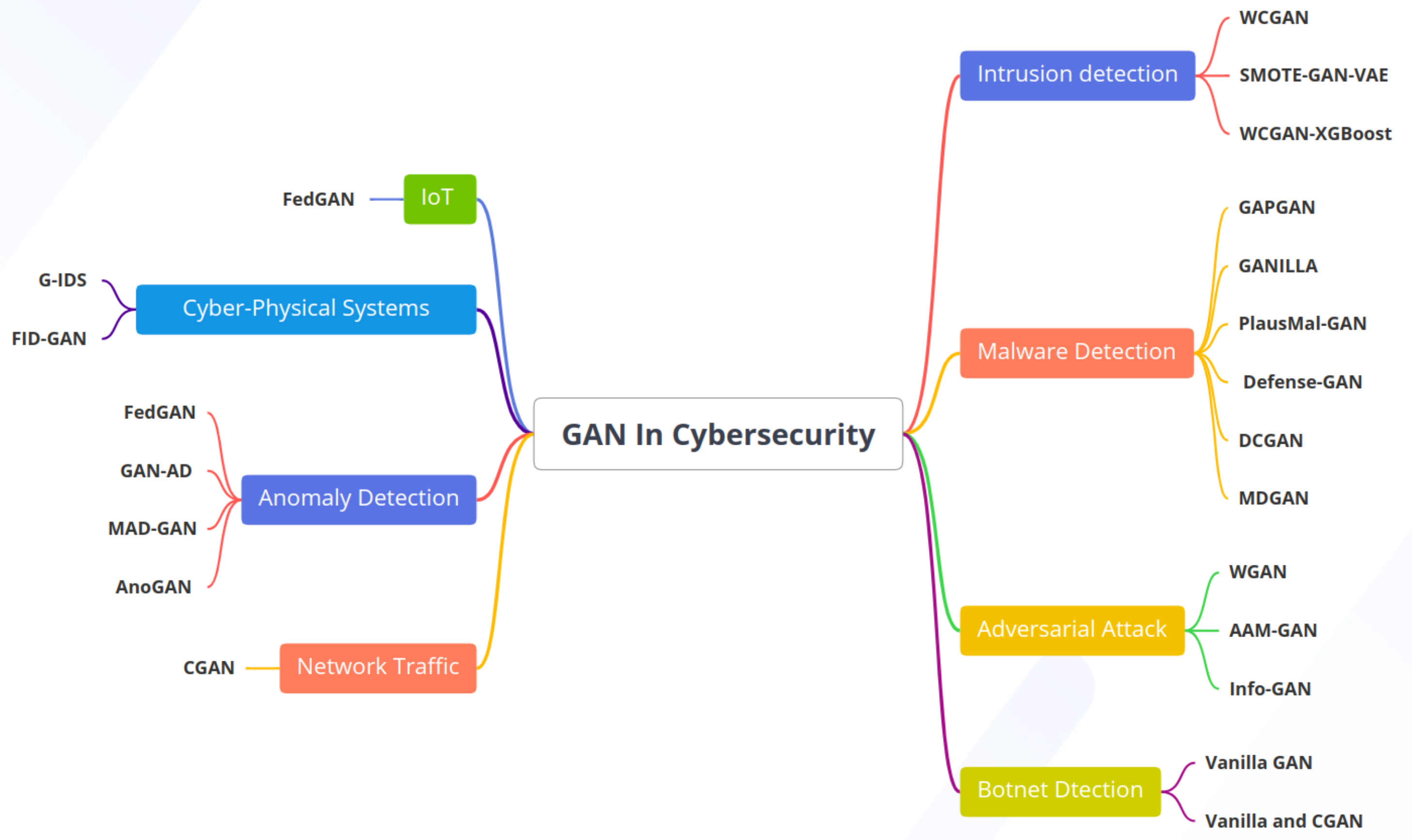} 
\caption{This taxonomy portrays the distribution of diverse GANs deployed in various cybersecurity domains.  }
\label{fig:GAN_Taxonomy}
 \end{figure*}

To improve understanding, a thorough summary of GAN properties and their uses in several cybersecurity fields is given in Table~\ref{table:gan-cybersecurity} and Figure~\ref{fig:GAN_Taxonomy}.

\begin{table*}[!t]
\centering
\caption{The table provides an overview of various Generative Adversarial Networks (GANs) employed in cybersecurity applications, highlighting their focus areas, learning methods, activation functions, and loss functions.}
\label{table:gan-cybersecurity}
\renewcommand{\arraystretch}{1.5} 
\begin{tabular}{p{2.5cm}p{2.5cm}p{3.5cm}p{3.5cm}p{2.5cm}}
\hline
\textbf{GAN} & \textbf{Focus Area} & \textbf{Learning Method} & \textbf{Activation Function} & \textbf{Loss Function} \\ \cline{1-5}
Vanilla GAN and CGAN~\cite{AS1_randhawa2021security} & Botnet & Adam Optimizer & Relu, Sigmoid & Binary cross-entropy (BCE) \\ 
WGAN and SMOTE-GAN-VAE~\cite{AS4_chui2023three} & NID & Grid search, genetic algorithm, particle swarm optimization & Relu, Sigmoid & Reconstruction loss function \\ 
WGAN~\cite{AS7_kim2022obfuscated} & Malware & Adam Optimizer & Leaky ReLU & N/A \\ 
PlausMal-GAN~\cite{AS8_won2022plausmal} & Zero-day malware& Adam Optimizer & Standard & Log loss function \\ 
WCGAN~\cite{AS11_kumar2023synthetic} & IDS & Keras tuner library & Relu, Sigmoid, nfolds & Standard \\ 
MDGAN~\cite{mazaed2022multifaceted} & Malware & N/A & N/A & Pixel-to-pixel conditional GAN \\ 
CGAN~\cite{yang2022network} & Network Security & N/A & Leaky ReLU, Sigmoid & N/A \\
\hline
\end{tabular}
\end{table*}

\section{Challenges}
Advanced GAN-based models have attracted significant attention in cybersecurity. Nonetheless, the integration of GANs into cybersecurity practices is confronted with many challenges. This section delves into the obstacles encountered in cybersecurity, encompassing the intricacies of safeguarding digital assets and the nuanced implementation of GANs. These obstacles range from the scarcity of diverse and quality datasets to the cat-and-mouse game with adversarial entities. Researchers and practitioners also face implementation challenges such as training instability, mode collapse, and the important task of harmonizing GANs with existing security frameworks.

\subsection{Cybersecurity Challenges}
In this context, we take a closer look at some of the state-of-the-art papers and provide easy-to-understand explanations for the challenges in cybersecurity.

\subsubsection{Adversarial Attacks and Adversarial Example Generation Challenges}
Radhawa et al.~\cite{AS1_randhawa2021security} propose a method to reduce the adverse effects of adversarial evasion attacks. These attacks are altered input samples designed to deceive the classifier. These attacks can compromise the accuracy of botnet detectors, creating significant security challenges. The research also focuses on improving recall, which is the ability of the classifier to identify botnet traffic from the dataset correctly. However, not all classifiers and datasets showed significant improvement in recall, which means that further investigation and improvement are necessary in this area. 

The research also addresses the problem of dataset imbalance and obsolescence, which can affect the effectiveness of botnet detectors. To keep these detectors up-to-date and effective, it is necessary to add new and unseen attacks to the datasets continuously. Generating large amounts of botnet data that have not been seen before can be difficult, but synthetic oversampling techniques like the use of GANs can help overcome this challenge. 

Moreover, the research also highlights the need to explore the behaviors and landscapes of modern botnets. This involves introducing new traffic features to differentiate botnets from normal traffic. Modern GANs can also be used to generate better-quality adversarial examples and improve the overall efficacy of botnet detection. 

These challenges underscore the complexity of addressing adversarial evasion attacks, dataset imbalance, and the need for continuous research and improvement in the field of botnet detection.

Zhang et al.~\cite{AS3_zhang2020brute} have highlighted the challenges that machine learning-based systems face in the realm of cybersecurity due to the emergence of adversarial examples (AEs). Adversarial examples are maliciously crafted noises added to original inputs with the aim of making target classifiers misclassify, thereby posing new threats to the security-critical applications in cybersecurity. The research points out that there is a lack of relevant studies in the domain of cybersecurity regarding AEs. Although some pioneering works have highlighted the vulnerability of deep neural networks to AEs, more attention needs to be paid to AEs in cybersecurity to address the new threats they pose to machine learning-based systems. The current research on AEs mainly focuses on computer vision, with only a few studies on AEs in cybersecurity. This highlights the need for further research to evaluate better the robustness of machine learning classifiers in cybersecurity against AEs. To summarize, the research underscores the challenges that machine learning-based systems face in cybersecurity due to the emergence of AEs. It stresses the need for more research in this domain to address the new threats posed by AEs and ensure the robustness of ML classifiers in cybersecurity.
     
In a different context, Schneider et al.~\cite{t4_schneider2023dual} focus on developing adversarial samples that can deceive both humans and machine learning models, highlighting a significant cybersecurity challenge. Adversarial attacks exploit vulnerabilities in ML systems, presenting a potential risk of incorrect outcomes and causing security breaches~\cite{zhuang2023robust}. This emphasizes the importance of developing robust cybersecurity measures to protect ML systems against adversarial attacks, which are becoming an increasing concern in the field.
     
Lucas et al.~\cite{t12_lucas2023adversarial} delve into the problem of adversarial examples, which are malicious binaries that are designed to dodge detection. These examples are modified in such a way that their functionality is preserved while they deceive the malware classifiers. The main challenge lies in developing robust models that can effectively detect and classify these adversarial examples. The study examines the effectiveness of adversarial training methods against state-of-the-art attacks. These attacks manipulate the bytes in non-executable parts of a binary to evade detection. The challenge is to create defenses that can reduce the impact of these attacks and improve the robustness of malware classifiers.

Defense-GAN,~\cite{laykaviriyakul2023collaborative}, a framework for protecting classifiers against adversarial attacks using generative models, has been used to tackle cybersecurity challenges related to adversarial attacks on deep learning models. Adversarial attacks are maliciously crafted perturbations that can fool pre-trained models, making them misclassify inputs. Defense-GAN has shown its effectiveness in countering both black-box and white-box attacks on datasets such as MNIST and F-MNIST. By incorporating GANs into the training process, Defense-GAN aims to improve the robustness of DL models against adversarial attacks in the field of cybersecurity.

Ferrag et al.~\cite{t9_ferrag2021federated} emphasize various challenges associated with implementing federated DL for cybersecurity in IoT applications. The paper highlights network-related issues like bandwidth limitations, interference, and noise that can hinder the development of effective federated learning systems. It stresses the complexity of managing data distributed across multiple parties in federated learning, particularly the challenge of mitigating bias and countering vulnerabilities like poisoning attacks and adversarial attacks, all while ensuring the practicality of deployment on resource-constrained IoT devices. The paper also emphasizes the need to design specific federated learning frameworks tailored to IoT applications, taking into account the unique characteristics of the IoT infrastructure. Lastly, the paper addresses the critical issue of privacy leakage attacks, which can occur when parameters of locally updated models on IoT devices reveal sensitive information. All these challenges collectively underscore the intricate and multifaceted nature of applying federated learning to enhance IoT cybersecurity.

\subsubsection{Interrpretibility of Deep Learning}

Ge et al.~\cite{t8_ge2023explainable} elucidates critical challenges permeating the realm of cybersecurity analysis. At the forefront of these challenges is a discernible trade-off between the performance prowess of AI models and their interpretability. While leveraging deep learning techniques can significantly elevate the quality of results, their inherent lack of transparency erects a formidable barrier for human operators. Understanding and trusting the decision-making processes of AI become intricate tasks, casting shadows on the seamless integration of these advanced models into cybersecurity operations.

A second pivotal challenge highlighted in the paper revolves around user comprehension. Existing methods grapple with providing comprehensive evidentiary support for recognition decisions, leaving human operators in a quandary. The insufficiency of information makes it challenging for operators to fully grasp and instill confidence in the outputs generated by AI systems. Bridging this gap between complex AI decisions and human interpretability emerges as a crucial area for improvement in the cybersecurity landscape.

The work also underscores the overarching complexity of cybersecurity analysis, which spans diverse domains, including network security, threat intelligence, and attack detection. This complexity begets challenges across the entire spectrum, from data analysis to model development and the amalgamation of diverse techniques. Navigating this multifaceted landscape necessitates concerted efforts in research and development. The call to action emanating from the challenges is clear – a trajectory toward the conception and implementation of more interpretable and understandable AI solutions in the cybersecurity domain. As the threat landscape evolves, the imperative for transparent and comprehensible AI models becomes increasingly paramount, guiding the trajectory of future advancements in cybersecurity research and practice.

The study~\cite{U1_ling2023adversarial} focuses on the absence of a clear understanding of DL models used in malware detection, which makes it difficult to understand how the models make decisions and identify vulnerabilities that can be exploited by attackers. There is difficulty in generating adversarial examples that are both effective and realistic, as the generated samples need to maintain the format, executability, and maliciousness of the actual malware samples. The hardship of generating adversarial examples can evade detection by multiple malware detection systems, as different systems may use different features and detection methods. The challenge of generating adversarial examples that are transferable means that they can be effective against different models or systems. Effective adversarial defenses are difficult to develop, as they must be able to detect and mitigate adversarial attacks while maintaining the performance of the malware detection system. This is a method of dealing with the dynamic nature of malware, as attackers can constantly modify and evolve their malware to evade detection.

\subsubsection{Data Related Challenge}
The detection of network attacks in cybersecurity poses a challenge when it comes to imbalanced datasets. Currently, the datasets used as benchmarks for network intrusion detection are imbalanced, indicating a significant difference in the number of samples available for different types of attacks as noted by Chui et al.~\cite{AS4_chui2023three} and Lee et al.~\cite{t20_lee2021gan}. This imbalance can lead to biased results and hinder the performance of intrusion detection models. A three-stage algorithm was proposed in the research to generate data that mitigates the impact of imbalanced ratios in minority classes. The algorithm uses a synthetic minority over-sampling technique, a generative adversarial network, and a variational autoencoder to generate data of high quality. By addressing this challenge, the research aimed to improve the accuracy and effectiveness of network intrusion detection models~\cite{zhang2020multiscale, bian2022learnable}. 

Kar et al.~\cite{kar2022integrated} have mentioned numerous cybersecurity challenges associated with wide-area control of power systems. One of the challenges is false data injection (FDI) attacks, which involve corrupting data streams from Phasor Measurement Units (PMUs) by hacking into the wide-area communication network channels. Another challenge is denial-of-service (DoS) attacks, which temporarily deactivate a link to obstruct the PMU data from reaching its desired destination. These attacks require effective detection and mitigation methods to ensure the security and safety of power systems.

Randhawa et al.~\cite{t5_randhawa2022evasion} delve into the intricate challenges prevalent in cybersecurity datasets, notably tackling the pervasive issue of data imbalance. In many instances, these datasets exhibit a disproportionality, with an abundance of normal behavior samples compared to the relatively scarce anomaly samples. This inherent data imbalance poses a substantial challenge, as it can introduce decision bias in machine learning classifiers, potentially compromising the efficacy of intrusion detection systems. The study critically addresses this data scarcity problem, recognizing the pivotal role it plays in the accurate classification of cyber threats. 

Shahriar et al.~\cite{t6_shahriar2020g} also thoroughly addresses the various cybersecurity challenges faced by Cyber-Physical Systems (CPS). The study also emphasizes the problems related to imbalanced and missing data in the training of Intrusion Detection Systems (IDS) for CPS, which hinders the accuracy of these systems. Additionally, the complexity and emergence of unknown behavior in CPS architectures pose additional challenges, further complicating the task of modeling and securing these systems. Lastly, the paper highlights the crucial need for a wider and more diverse range of training data to enhance the accuracy and effectiveness of IDS models, especially in the context of emerging CPS technologies.

\subsubsection{Malware Detection Related challenges}

The multifaceted domain of cybersecurity with a primary focus on malware detection is a formidable challenge within the field mentioned by Kim et al.~\cite{AS7_kim2022obfuscated}. To tackle this issue, the study introduces an innovative approach employing both global and local features, surpassing conventional methods and attaining state-of-the-art performance. 
Huang et al.~\cite{huang2022android} found that relying solely on images to detect Android malware is insufficient and can lead to lower accuracy. Their study highlighted the need for better cybersecurity measures and suggested that quantum advantage could improve existing detection methods and overcome bottlenecks. However, implementing quantum computing techniques in this field is complex and challenging.

The study by Won et al.~\cite{AS8_won2022plausmal} takes a critical stance on the pressing issue of detecting and classifying zero-day malware. This category encompasses previously unknown and unanalyzed malware capable of exploiting vulnerabilities in computer systems. The traditional reliance on malware signatures by antivirus systems is scrutinized, revealing its inefficiency and ineffectiveness in the face of the rapidly evolving landscape of zero-day threats.

In a parallel pursuit, Alotaibi et al.~\cite{mazaed2022multifaceted} illuminate a multitude of cybersecurity challenges inherent in malware detection and protection within the Android operating system. The study underscores the significant security threat posed by phishing attacks, which cleverly deceive users into unwittingly installing malware through malicious URLs. Adding complexity to the threat landscape are intelligent malware-producing applications, contributing to the proliferation of diverse malware variants. The limitations of signature-based detection tools and the intricacies associated with static and dynamic analysis methods further compound the challenges in accurately detecting and classifying malware.

Liu et al.~\cite{t2_liu2020efficient} discuss several cybersecurity challenges. Firstly, they address the issue of the increasing number of malware, which poses a significant challenge to cybersecurity. Malware is diverse and complicated, making it challenging for intrusion detection systems to combat them effectively. Secondly, they discuss the problem of malware variants and concealment. Malware authors use techniques such as packing and obfuscation to create numerous variants that outpace signature-based detection methods. Additionally, they highlight the challenge of detecting and analyzing malicious traffic patterns initiated by these variants, especially on servers that store sensitive information. Traditional detection schemes are also vulnerable to adversarial samples, making the development of anomaly detection methods crucial. Finally, the challenge of limited data samples and the need to detect zero-day exploits has been addressed.

Moreover, wang et al.~\cite{t14_wang2022evilmodel} highlight the emerging threat of using neural network models as carriers of malware. Network security faces a considerable challenge due to the ability to switch a large number of parameters in standard neural network models with malware bytes or other types of information without any impact on the model's performance.

\subsubsection{Intrusion Detection Related Challenge}

Kumar et al.~\cite{AS11_kumar2023synthetic} tackle the significant challenge of accurately detecting a multitude of attack classes in intrusion detection systems (IDS) while emphasizing the critical need for representative samples within each attack class. Data imbalance issues often plague IDS datasets, resulting in biased intrusion detection models that favor majority classes. 
We observe that Soleymanzadeh et al.~\cite{soleymanzadeh2022stable} have tackled several cybersecurity challenges related to network intrusion detection. One of the challenges addressed in the paper is detecting multi-attack types, which requires a classification model that is both robust and accurate. In addition, the paper also addresses the challenge of stability in the training process for GANs.
The cybersecurity challenges posed by cyberattacks and intrusions in the context of IoTs and smart devices have been mentioned by Duy et al.~\cite{t1_duy2021digfupas}. In addition, it highlights the importance of having remedies that can fill in the gaps in the training data and offer automatic self-improvement and resilience enhancement while operating in real time.

\subsubsection{Network Security Related Challenge} 

Yang et al.~\cite{yang2022network} emphasized several cybersecurity challenges in network security. They underlined that the rapid development of computer technology and the expanding number of internet users had significantly increased network traffic, making detecting and responding to network threats more difficult. The emergence of new network threats, for instance, trojan horses, viruses, and phishing sites, further complicated the network security situation. Additionally, the expansion of the fifth-generation mobile communication network has resulted in explosive growth in network traffic, posing a severe challenge to network managers. These challenges necessitated the development of effective methods for network security situation element extraction and proactive defense against network threats.

In the study~\cite{t11_das2022fgan}, the authors bring to light the challenges of dealing with the constantly changing nature of computer and mobile networks. As the number of nodes and the volume of traffic increase, detecting anomalies and adapting to modern attacks become more complex. Accurately detecting network intrusions is also a challenge mentioned in the study, along with concerns about data privacy in intrusion detection systems. The suggested approach uses federated learning to tackle these problems. This enables the utilization of training data at local levels and the sharing of encrypted models, all while guaranteeing the protection of confidential information. The study also recognizes the challenge of coordinating model updates in large-scale networks and highlights the problem of model tampering, where attackers may try to modify the trained model.

Park et al.~\cite{t7_park2022enhanced} identifies several significant challenges in the realm of network security. These challenges include the proliferation of diversified access points in the context of evolving 5G technology and distributed networks, leading to an expanded attack surface that makes network systems more vulnerable to potential threats. Additionally, the paper underscores the increasing complexity and sophistication of cyber-attacks, coupled with a rising frequency of such attacks, which present substantial difficulties in detecting and preventing network threats. Moreover, it emphasizes the heightened importance of cybersecurity due to the growing vulnerability of network systems and a strong focus on enhancing network intrusion detection as a preventive measure. Ultimately, the study emphasizes the critical nature of addressing these challenges in the ongoing efforts to safeguard network systems from various threats and intrusions.

Botnets are becoming increasingly sophisticated and intelligent, making it challenging to detect their activities. Their methods are constantly changing, utilizing cutting-edge concepts and technologies to evade discovery, which presents a major obstacle in the realm of cybersecurity. Limitations exist in current network-based detection methods for botnets. The study conducted by Yin et al.~\cite{yin2018enhancing} focused solely on certain characteristics of network flows, which are insufficient in fully describing the anomalous activities of botnets.

\subsubsection{AI Generated Attack} 

Gupta et al.~\cite{t10_gupta2023chatgpt} address the obstacles involved in circumventing the ethical and privacy protections of artificial intelligence models such as ChatGPT. The work highlights the potential cyber-attacks that can be launched using ChatGPT and explores the use of Generative AI (GenAI) in cyber offenses. Moreover, it emphasizes the importance of addressing the security implications of AI-powered attacks to protect user privacy and ethical boundaries. Additionally, the research paper demonstrates how these models can be attacked using reverse psychology and jailbreak techniques.

\subsubsection{Vulnerability of conventional cryptographic functions }
Hao et al.~\cite{hao2021asymmetric} brought to the forefront several profound challenges that significantly impact the security landscape of cyber-physical systems (CPS) and IoTs. Their insightful analysis unveils a multifaceted spectrum of issues, highlighting the intricate intersection of physical and cyberspace threats that pose complex challenges in safeguarding these interconnected systems. These challenges span from covert surveillance and unauthorized data leakage to a myriad of sophisticated cyber attacks.

A particularly noteworthy challenge elucidated by the authors pertains to the vulnerability of conventional cryptographic functions, including widely utilized ones such as RSA and DES, in the face of quantum computing advancements. The looming threat of quantum computing renders digital signatures and encryption schemes susceptible to interception and modification, necessitating a reevaluation of cryptographic strategies to ensure resilience against emerging technological paradigms.

Furthermore, the authors have highlighted the need for a delicate equilibrium between maintaining high efficiency and ensuring security in the IoT industry. Existing signature schemes may prove inadequate in meeting these dual demands, and the authors caution against potential shortcomings in employing symmetric encryption in conjunction with Generative Adversarial Networks (GANs). The nuanced complexities in this arena are particularly apparent in the realm of key management and exchange security, where innovative strategies are essential to address the evolving landscape of IoT security.

\subsection{Implementation Challenges for Cybersecurity}
In this section, we'll be exploring the difficulties faced by researchers when building GAN models with a focus on cybersecurity. We believe this information will make the researchers aware of designing the model in this domain. 

Imbalanced and limited data pose significant challenges in the context of GANs. These issues can impact the training and performance of GAN models, particularly in scenarios where data scarcity or imbalances between classes are prevalent. Ensuring balanced and robust data generation by GANs is imperative for effective model training. In the study~\cite{AS1_randhawa2021security}, GANs encounter challenges mitigating adversarial attacks and handling data imbalance in botnet datasets. The framework for malware detection~\cite{mazaed2022multifaceted} highlights potential challenges with GANs related to limited training data availability. Additionally, the reliance on expert knowledge for handcrafted feature extraction may hinder accessibility for non-experts. GANs can contribute to data diversity, but their performance hinges on the availability of high-quality data. In intrusion detection systems, GANs can address the challenge of imbalanced data~\cite{soleymanzadeh2022stable} by generating synthetic samples for minority classes and balancing the dataset. Moreover, the utilization of GANs, such as Conditional GANs (CGANs), to address data imbalance in network traffic data~\cite{yang2022network} requires meticulous parameter tuning and validation for the effective generation of minority class samples. Last but not least, the context of LSTM-GAN methods~\cite{kar2022integrated} for power system security introduces challenges related to data sampling. Inadequate sampling during training can lead to inaccurate identification and control actions.

Combating the never-ending competition between attackers and defenders in detecting Windows PE malware is a challenging task. Incomplete evaluations and inherent biases hinder the comprehensive assessment of these methods. In parallel domains like adversarial attacks in texts, images, and graphs, the availability of numerous open-source toolboxes and platforms~\cite{U25_zeng2020openattack, U26_papernot2016technical, U27_ling2019deepsec, U28_li2020deeprobust, li2024mapping} has significantly advanced research. However, in the specific context of Windows PE malware detection, a notable gap exists in the provision of such comprehensive resources. The mentioned challenges in quantifying the efficacy of adversarial attacks and defenses find a potential ally in GANs. Leveraging their generative capabilities, GANs facilitate the creation of a diverse array of adversarial samples, enriching the evaluation process for detection models. This capability proves especially vital in understanding the nuanced strengths and limitations of existing defense mechanisms. Proposing the integration of GAN-based toolboxes specific to this domain not only addresses this gap but also revolutionizes the evaluation landscape. Such platforms, enriched by GANs, not only empower researchers to scrutinize the effectiveness of current methods but also serve as a springboard for the development of novel attacks and defenses. The collaborative ethos seen in other fields can be transposed to Windows PE malware detection through the incorporation of GANs into evaluation frameworks.

The computational complexity of Generative Adversarial Networks (GANs) is a crucial aspect that influences their efficiency and scalability. GANs can assist in generating diverse malware samples to tackle the challenge of limited availability~\cite{AS7_kim2022obfuscated}. However, the introduction of computational complexity in GAN-based approaches necessitates substantial computational resources and time. GANs, employed for robustness against adversarial attacks in~\cite{AS9_reilly2023robustness}, encounter resource constraints due to the computational intensity of training deep learning models. Data preparation challenges, crucial for GAN-based data generation, may also arise.

The study~\cite{AS8_won2022plausmal} underscores the role of GANs in addressing challenges related to adversarial attacks and zero-day malware detection. Optimizing GAN models for robustness against adversarial examples is critical, and GANs can augment malware images to enhance training dataset diversity. GANs, as utilized to generate malware variants to evade known AV detectors~\cite{babaagba2023evolutionary}, require fine-tuning to achieve high adversarial accuracy. Also, GANs generate mutants capable of fooling detectors, presenting a challenge in classification. Moreover, GANs relying on Phasor Measurement Unit (PMU) data for detecting and mitigating cyber-attacks~\cite{kar2022integrated} may face challenges if the PMU data's availability or accuracy is compromised, directly impacting system performance.

\begin{table*}[H]
    \centering
    \caption{Cybersecurity Challenges Overview}
    \begin{tabular}{|p{3cm}|p{5cm}|p{8cm}|}
        \hline
        \textbf{Category} & \textbf{Challenges} & \textbf{Key Points} \\ \hline
        Adversarial Attacks & Adversarial evasion attacks, dataset imbalance, continuous research & 
        - Reducing effects of adversarial evasion attacks \cite{t5_randhawa2022evasion}  \newline
        - Dataset imbalance and obsolescence (GANs for synthetic data) \newline
        - Modern botnets behavior and traffic features \cite{t4_schneider2023dual} \cite{t12_lucas2023adversarial} \newline
        - Adversarial examples in cybersecurity \cite(Zhang et al.) \newline
        - Robust models against adversarial samples (Defense-GAN)\\ \hline
        
        Interpretability of Deep Learning & Trade-off between performance and interpretability, user comprehension, complexity of cybersecurity analysis & 
        - Lack of transparency in AI models \cite{t3_ge2023explainable} \newline
        - Challenges in user comprehension and evidence support \newline
        - Complexity across various cybersecurity domains \newline
        - Necessity for interpretable AI solutions \cite{U27_ling2019deepsec} \\ \hline
        
        Data Related Challenges & Imbalanced datasets, false data injection, data scarcity & 
        - Imbalanced network intrusion datasets \cite{AS4_chui2023three}\cite{AS4_chui2023three} \newline
        - FDI and DoS attacks in power systems \cite{kar2022integrated} \newline
        - Data imbalance in cybersecurity datasets \cite{AS1_randhawa2021security}\cite{t6_shahriar2020g} \\ \hline
        
        Malware Detection & Zero-day malware, obfuscated malware, AI-driven malware & 
        - Detection of zero-day malware \cite{AS8_won2022plausmal} \newline
        - Android malware detection challenges \cite{huang2022android}\cite{mazaed2022multifaceted} \newline
        - Increasing number of malware variants \cite{liu2019machine} \cite{wang2024balanced} (Liu et al., Wang et al.) \\ \hline
        
        Intrusion Detection & Multi-attack detection, real-time operation & 
        - Accurate detection of various attack classes \cite{AS11_kumar2023synthetic} (Kumar et al.) \newline
        - Stability in GAN training \cite{soleymanzadeh2022stable}(Soleymanzadeh et al.) \newline
        - Gaps in training data for IoTs and smart devices \cite{t1_duy2021digfupas} (Duy et al.) \\ \hline
        
        Network Security & Increased network traffic, new network threats, privacy concerns & 
        - Complexity in detecting network threats \cite{yang2022network} (Yang et al.) \newline
        - Adapting to modern attacks and privacy in IDS \cite{t11_das2022fgan} \cite{t7_park2022enhanced}(Das et al., Park et al.) \newline
        - Detection limitations in botnet activities \cite{yin2018enhancing} (Yin et al.) \\ \hline
        
        AI Generated Attacks & Ethical and privacy protections in AI models & 
        - Cyber-attacks using AI like ChatGPT (Gupta et al.) \newline
        - Generative AI in cyber offenses \newline
        - Addressing AI-powered attack implications \\ \hline
        
        Vulnerability of Cryptographic Functions & Impact of quantum computing on cryptographic functions & 
        - Challenges in safeguarding CPS and IoTs \cite{hao2021asymmetric} (Hao et al.) \newline
        - Vulnerability of conventional cryptographic functions (RSA, DES) \newline
        - Balancing efficiency and security in IoT \\ \hline
        
        Implementation Challenges for Cybersecurity & Imbalanced data, computational complexity, incomplete evaluations & 
        - Data generation issues in GANs \cite{AS1_randhawa2021security} \cite{AS7_kim2022obfuscated} (Randhawa et al., Kim et al.) \newline
        - Computational resource constraints \newline
        - Optimizing GANs for adversarial robustness \cite{AS9_reilly2023robustness} (Reilly et al.) \newline
        - GANs for malware variant generation and classification challenges \cite{babaagba2023evolutionary} (Babaagba et al.) \\ \hline
    \end{tabular}
    \label{tab:cybersecurity_challenges}
\end{table*}

\section{Use Cases and Applications}
This section provides an overview of how GAN can be applied to protect various fields. Each part investigates a specific application of GAN, detailing its intricacies, challenges, and potential solutions using cutting-edge technology. Additionally, Table~\ref{table:summary_of_use_cases} and \ref{table:summary_of_use_cases_2} in this section are valuable resources for compiling essential insights and discoveries from these diverse applications.

\begin{table*}[t]
\centering
\caption{Summary of use cases and applications regarding GAN }
\label{table:summary_of_use_cases}
\begin{tabular}{|l|l|l|} 
\hline
    \textbf{Implementation Source} & \textbf{Application Domain} & \textbf{Main Idea} \\                                                          
\hline
Randhawa et al.~\cite{AS1_randhawa2021security}                                & BotNet                                                                                            & \begin{tabular}[c]{@{}l@{}}GANs improve botnet detection by generating high-quality \\ traffic samples, addressing challenges like adversarial evasion \\ attacks. They also promise to enhance autonomous IDS \\ through proactive training against novel evasion samples \\ using deep reinforcement learning.\end{tabular}                                                                                      \\ 
\hline
Alotaibi et al.~\cite{mazaed2022multifaceted}                                    & Mobile Trespass                                                                                   & \begin{tabular}[c]{@{}l@{}}Developing a robust malware detection framework for mobile \\ devices is addressed by proposing a novel approach based on \\ DCNN and attention mechanisms, combining visual and  \\ semantic feature extraction.\end{tabular}                                                                                                                                                           \\ 
\hline
Yang et al.~\cite{yang2022network}                                       & \multirow{5}{*}{Network Trespass}                                                                 & \begin{tabular}[c]{@{}l@{}}Combining CGAN and Transformer models, Yang et al.'s \\ approach to network security enhances the detection of \\ minority samples and overall accuracy in situation element \\ extraction, contributing to more effective identification and \\ mitigation of network threats.\end{tabular}                                                                                            \\ 
\cline{1-1}\cline{3-3}
Soleymanzadeh et al.~\cite{soleymanzadeh2022stable}                            &                                                                                                   & \begin{tabular}[c]{@{}l@{}}This approach aids in identifying and preventing network \\ intrusions, including DoS attacks, unauthorized access, \\ and probing attacks, by modeling complex distributions \\ of high-dimensional data for effective anomaly detection.\end{tabular}                                                                                                                                  \\ 
\cline{1-1}\cline{3-3}
Duy et al.~\cite{t1_duy2021digfupas}                                       &                                                                                                   & \begin{tabular}[c]{@{}l@{}}DIGFuPAS, a framework integrated into a network security \\ automation pipeline for commercial applications. This \\ incorporation enables the automation of generating \\ adversarial samples and evaluating IDS performance in \\ SDN-enabled networks.\end{tabular}                                                                                                                  \\ 
\cline{1-1}\cline{3-3}
Chalé et al.~\cite{U7_chale2022generating}                                        &                                                                                                   & \begin{tabular}[c]{@{}l@{}}This innovative approach aims to enhance the accuracy and  \\ effectiveness of machine learning classifiers, thereby improving  \\ the military's capability to detect and respond to cyber threats.\end{tabular}                                                                                                                                                                      \\ 
\cline{1-1}\cline{3-3}
Wang et al.~\cite{t14_wang2022evilmodel}                                        &                                                                                                   & \begin{tabular}[c]{@{}l@{}}This discovery underscores the emerging threat of embedding \\  malware in neural network models, emphasizing the need for \\ heightened awareness and preparedness among security  \\ researchers to address this significant threat to network security.\end{tabular}                                                                                                                    \\ 
\hline
Kim et al.~\cite{AS7_kim2022obfuscated}                                         & \multirow{4}{*}{\begin{tabular}[c]{@{}l@{}}Malware Detection\\ (Obfuscated malware)\end{tabular}} & \begin{tabular}[c]{@{}l@{}}This approach, which transforms malware into visual \\ representations and extracts local characteristics from binary  \\ code sequences, holds practical implications for computer \\ security,  especially in malware detection and mitigation, \\ reinforcing communication network security against cyber threats.\end{tabular}                                                       \\ 
\cline{1-1}\cline{3-3}
Yang et al.~\cite{yang2022network}                                        &                                                                                                   & \begin{tabular}[c]{@{}l@{}}This approach, capable of generating variations and camouflage \\ of malware, not only effectively mitigates new threats in  \\  real-time detection scenarios but also holds potential applications  \\ in diverse domains,  including sound and speech processing, \\  hardware security, artificial intelligence  \\ security, and emerging technologies.\end{tabular}                     \\ 
\cline{1-1}\cline{3-3}
Liu et al.~\cite{t2_liu2020efficient}                                        &                                                                                                   & \begin{tabular}[c]{@{}l@{}}The objective is to improve the protection of specific targets by  \\ accurately detecting and mitigating the threat posed by \\ malware attacks.\end{tabular}                                                                                                                                                                                                                           \\ 
\cline{1-1}\cline{3-3}
Lucas et al.~\cite{t12_lucas2023adversarial}                                       &                                                                                                   & \begin{tabular}[c]{@{}l@{}}The focus is on developing defenses against sophisticated evasion  \\ methods used by attackers to deceive malware classifiers, aiming \\ to create  raw-binary malware classifiers resistant to  \\ state-of-the-art  evasion techniques through experimentation \\ with adversarial training.\end{tabular}                                                                                \\ 
\hline
Won et al.~\cite{AS8_won2022plausmal}                                        & \multirow{2}{*}{\begin{tabular}[c]{@{}l@{}}Malware Detection\\ (Zeroday malware)\end{tabular}}    & \begin{tabular}[c]{@{}l@{}}Exploring explainable AI techniques to enhance understanding of  \\ zero-day malware characteristics, with broad applicability  \\  in scenarios critical for upholding computer security.\end{tabular}                                                                                                                                                                                 \\ 
\cline{1-1}\cline{3-3}
Reilly et al.~\cite{AS9_reilly2023robustness}                                    &                                                                                                   & \begin{tabular}[c]{@{}l@{}}Utilizing deep learning algorithms for zero-day malware  \\ classification and considers the potential improvement in model \\ robustness through training on GAN generated samples.\end{tabular}                                                                                                                                                                                     \\ 
\hline
Babaagba et al.~\cite{babaagba2023evolutionary}                                    & \begin{tabular}[c]{@{}l@{}}Malware Detection\\ (Metamorphic malware)\end{tabular}                 & \begin{tabular}[c]{@{}l@{}}The primary objective is to generate malware samples capable of  \\ deceiving detection systems by crafting samples that can fool  \\ antivirus engines and machine learning detectors into classifying  \\ them as harmless.\end{tabular}                                                                                                                                                 \\ 

\hline
\end{tabular}
\end{table*}

\begin{table*}[t]
\centering
\caption{Summary of use cases and applications regarding GAN (Cont.) }
\label{table:summary_of_use_cases_2}
\begin{tabular}{|l|l|l|} 
\hline
    \textbf{Implementation Source} & \textbf{Application Domain} & \textbf{Main Idea} \\                                                          
\hline

Rathore et al.~\cite{U5_rathore2023adversarial}                                    & \multirow{2}{*}{Attacks on Android}                                                               & \begin{tabular}[c]{@{}l@{}}Developing adversary-aware Android malware detection models, \\ investigating their adversarial robustness and emphasizing the  \\ uniquechallenges posed by  Android malware in the  \\ cybersecurity landscape.\end{tabular}                                                                                                                                                            \\ 
\cline{1-1}\cline{3-3}
Wu et al.~\cite{AS6_wu2020network}                                          &                                                                                                   & \begin{tabular}[c]{@{}l@{}}An efficient black-box adversarial attack framework for Android \\  devices, leveraging an attention model and optimization \\  techniques to generate scalable adversarial \\ examples, enhancing real-world applicability\end{tabular}                                                                                                                                                  \\ 
\hline
Moti et al.~\cite{U9_moti2021generative}                                      & \multirow{7}{*}{Tresspass on IoT}                                                                 & \begin{tabular}[c]{@{}l@{}}Enhanced accuracy and stability in detecting malware in IoT  \\ devices, demonstrating promising results on standard IoT \\ and Windows malware datasets.\end{tabular}                                                                                                                                                                                                                   \\ 
\cline{1-1}\cline{3-3}
Ferrag et al.~\cite{t9_ferrag2021federated}                                      &                                                                                                   & \begin{tabular}[c]{@{}l@{}}Contributed a novel generative adversarial network and  \\ Transformer-basedmodel to address cybersecurity challenges  \\  in 6G-enabled  IoT networks, utilizing generative AI for real-time  \\  detection and prevention of cyber attacks in this complex  \\  and heterogeneous environment.\end{tabular}                                                                                  \\ 
\cline{1-1}\cline{3-3}
Bakhsh et al.~\cite{U12_bakhsh2023enhancing}                                     &                                                                                                   & \begin{tabular}[c]{@{}l@{}}Employed a traffic capture mechanism to collect data flow from  \\ IoT devices, simulating operations within an  \\  IoT-enabled Smart Home.\end{tabular}                                                                                                                                                                                                                                  \\ 
\cline{1-1}\cline{3-3}

Gupta et al.~\cite{t10_gupta2023chatgpt}                                      &                                                                                                   & \begin{tabular}[c]{@{}l@{}}Developed innovative security mechanisms, models, and   \\architectures for next-generation smart cars, intelligent transportation \\ systems, and smart farming.\end{tabular}                                                                                                                                                                                                           \\ 
\hline
Kumar et al.~\cite{AS11_kumar2023synthetic}                                      & \multirow{4}{*}{Intrusion Detection}                                                              & \begin{tabular}[c]{@{}l@{}}Highlighted the model's versatility in various domains, such as \\ medical diagnostics and financial anomaly detection, \\ demonstrating its potential to improve attack detection \\performance using GANs.\end{tabular}                                                                                                                                                                \\ 
\cline{1-1}\cline{3-3}
Duy et al.~\cite{t1_duy2021digfupas}                                        &                                                                                                   & \begin{tabular}[c]{@{}l@{}}Evaluated and enhance the robustness of IDS by generating \\ high-quality attack  traffic of different types.\end{tabular}                                                                                                                                                                                                                                                            \\ 
\cline{1-1}\cline{3-3}
Shahriar et al.~\cite{t6_shahriar2020g}                                   &                                                                                                   & \begin{tabular}[c]{@{}l@{}}Proposed framework has implications for strengthening intrusion \\ detection and  enhancing security in CPS, especially in emerging \\ domains with limited data availability.\end{tabular}                                                                                                                                                                                            \\ 
\cline{1-1}\cline{3-3}
Park et al.~\cite{t7_park2022enhanced}                                       &                                                                                                   & \begin{tabular}[c]{@{}l@{}}Resolved the data imbalance problem and improved the classification  \\ performance to evaluate on various data sets, including benchmark  \\ data sets, an IoT data set, and a real data set.\end{tabular}                                                                                                                                                                            \\ 
\hline
Dirgantoro et al.~\cite{U10_dirgantoro2020generative}                                & Face Recognition                                                                                  & \begin{tabular}[c]{@{}l@{}}Utilized edge computing to reduce latency and a low static  \\ difficulty blockchain to facilitate transactions.\end{tabular}                                                                                                                                                                                                                                                         \\ 
\hline
Kar et al.~\cite{kar2022integrated}                                        & \begin{tabular}[c]{@{}l@{}}False Data Injection and \\ Denial-of-Services\end{tabular}            & \begin{tabular}[c]{@{}l@{}}Developed a deep learning method based on GANs to identify  \\ and mitigate false data injection and denial-of-service \\ attacks in power systems' wide-area damping control loops\end{tabular}                                                                                                                                                                                      \\ 
\hline
Ge et al.~\cite{t3_ge2023explainable}                                         & \begin{tabular}[c]{@{}l@{}}Cyber Behavior \\ Characterization\end{tabular}                        & \begin{tabular}[c]{@{}l@{}}Locating and decomposing CTI texts into specific entities and  \\ forming an attack behavior graph, analysts can determine which  \\ attack techniques should be focused on and understand the next \\ steps of the attack.\end{tabular}                                                                                                                                                  \\
\hline
\end{tabular}
\end{table*}

\subsection{Botnet Detection}
Randhawa et al.~\cite{AS1_randhawa2021security} demonstrates that Generative Adversarial Networks (GANs) can be effectively used in computer vision-based applications as well as in the cybersecurity domain. Specifically, GANs can be utilized to address challenges related to adversarial evasion attacks in botnet detection. By generating high-quality traffic samples, GANs can enhance the performance of botnet detectors, improve security hardening, and mitigate false positives. Additionally, GANs can be used to balance datasets and introduce new traffic features to differentiate botnets from normal traffic. The study also suggests the potential use of GANs in autonomous IDS for proactive training against novel evasion samples using deep reinforcement learning. Another proposed framework~\cite{yin2018enhancing}, Bot-GAN, aims to enhance botnet detection models. A generator is used to produce fabricated samples, which helps augment the labeled data and enhance the performance and generalization of the initial model. After the augmentation, the improved model is evaluated to identify unknown botnets and other entities that make use of encryption or proprietary protocols. It showed promising results in improving the detection performance, decreasing the false positive rate, and providing a practical approach for detecting unknown botnets, which leads to the suitability for anomaly detection in network traffic.

\subsection{Mobile Trespass}
Detecting malware in mobile devices is challenging due to the dynamic and evolving nature of the malicious software~\cite{feng2020performance}. The existing malware detection frameworks depend on static or dynamic analysis of the application files, which have limitations in accuracy, efficiency, and scalability~\cite{amro2018malware, shabtai2010malware}. A novel malware detection framework established on deep neural networks (DCNN) and attention mechanisms are proposed in this work~\cite{mazaed2022multifaceted}. This framework extracts visual and semantic features from the application files using DCNN and attention mechanism, respectively. This innovative approach incorporates the power of deep learning, generative adversarial networks (GANs), and sophisticated feature processing techniques to detect malware in mobile devices. 

\subsection{Network Trespass}
Yang et al.~\cite{yang2022network} focused on network security concerns using a combination of Conditional Generative Adversarial Network (CGAN) and Transformer models. The CGAN and Transformer model can detect and classify network traffic. By combining these models, the proposed approach improves the detection pace of minority instances and the accuracy of comprehensive situation component extraction. It can help identify and mitigate network threats more effectively. Another proposed architecture~\cite{soleymanzadeh2022stable} is employed to enhance network security by detecting and categorizing various types of malicious activities. It can help identify and prevent network intrusions, such as unauthorized access, probing attacks, and denial-of-service (DoS). Using generative adversarial networks (GANs) in the proposed architecture allows for modeling complex distributions of high-dimensional data. It can be particularly helpful for detecting unusual occurrences, where the system can identify abnormal patterns or behaviors that deviate from the expected network activity.

Duy et al.~\cite{t1_duy2021digfupas} proposed a framework designed to be integrated into the pipeline, and it aimed to improve network security for businesses. By incorporating DIGFuPAS into the pipeline, it becomes possible to automate the process of generating adversarial samples and evaluating the performance of IDS in SDN-enabled networks.

The problem stated in~\cite{U7_chale2022generating} is that the availability of realistic cyber data for training and evaluation limits current machine learning classifiers for network intrusion detection systems. The solution proposed in this paper file is to use generative models, specifically GANs and VAEs, to develop realistic cyber data that can be exploited to train and assess machine learning classifiers for network IDS. This approach can help improve the accuracy and effectiveness of these classifiers, ultimately enhancing the military's capability to detect and respond to cyber threats.

Wnag et al.~\cite{t14_wang2022evilmodel} explore using neural network models as carriers for malware. It demonstrated that many parameters in formal neural network prototypes can be restored with malware bytes without affecting the model's performance. This finding has significant implications for network security and highlights the need for security researchers to be prepared for this emerging threat. The work also emphasizes that embedding malware in neural network models poses a significant threat to network security. 

\subsection{Malware Detection}
\subsubsection{ Obfuscated malware Detection}
This research~\cite{AS7_kim2022obfuscated} paper delved into applying a deep generative model for detecting obfuscated malware, presenting a novel approach that amalgamates global and local features. This proposed method exhibits notable efficacy in identifying malware variants by transforming malware into visual representations to capture global attributes and utilizing binary code sequences to extract local characteristics. Its practical implications extend to the realm of computer security, notably in the realm of malware detection and mitigation. In sum, integrating deep generative models into obfuscated malware detection can fortify communication network security and bolster defenses against cyber threats. Yang et al.~\cite{yang2022network} proposed a method (Visual-AT) that can enhance malware detection systems' accuracy and robustness. Exploiting Autoencoder and Adversarial Training (AT) techniques provide a suitable model regularization for better performance. Visual-AT can generate variations and camouflage of malware, effectively mitigating new and potential threats in real-time detection scenarios. This method can be extended to new scenarios, such as hierarchical sets of labels (e.g., Animal-Dog-Poodle). It opens up possibilities for applying the method in various domains, including sound and speech processing for recognition impairment. It also has the potential for an expansive field of machine learning and computer safety applications. It can be utilized in different fields beyond malware detection, contributing to advancements in areas like hardware security, artificial intelligence security, and emerging technologies.

Liu et al.~\cite{t2_liu2020efficient} focused on detecting malicious traffic generated by malware variants. TrafficGAN is a system that has been proposed to generate various adversarial traffic patterns using a conditional GAN. The system aims to enhance the protection of specific targets by accurately detecting and mitigating the threat of malware attacks. 

Lucas et al.~\cite{t12_lucas2023adversarial} aimed to enhance the robustness of malware classifiers by using adversarial training strategies. It can help strengthen the precision and efficiency of malware detection systems, which can be beneficial in improving cybersecurity measures. This research focuses on developing defenses against sophisticated evasion methods used by attackers to fool malware classifiers. This study intended to create raw-binary malware classifiers that resist state-of-the-art evasion methods by experimenting with different adversarial training techniques.

\subsubsection{Zeroday malware Detection}
The PlausMal-GAN framework~\cite{AS8_won2022plausmal} leveraged generative adversarial networks (GANs) to identify analogous zero-day malware, focusing on the variants of malware with subtle distinctions from known samples. Training a discriminator with generated malware images achieves consistent and high accuracy in identifying such threats without relying on inefficient signature analysis. This paper also outlined potential future research avenues, including broadening the malware datasets, optimizing the GAN model, and exploring explainable AI techniques to enhance the understanding of zero-day malware characteristics. The framework that has been suggested can be used in situations where it is essential to identify and categorize comparable zero-day malware in order to maintain computer security. Another research~\cite{AS9_reilly2023robustness} focused on applying deep learning algorithms for zero-day malware classification and It enhances the strength of models by training them on samples generated by GANs. The study also explored the space-filling curve method and its computational and space-saving benefits, suggesting the need for further investigation into enhancing this process.
\subsubsection{ Metamorphic malware }
Metamorphic malware is a growing concern for computer systems and networks since it can easily slip past detection by constantly changing its form. In a recent paper~\cite{babaagba2023evolutionary}, a new approach was proposed to address this issue. The approach employs an Evolutionary-inspired technique, using an Evolutionary Algorithm as the generator. The primary aim is to generate malware samples that can deceive detection systems into classifying them as harmless. It is done by crafting samples that are capable of fooling antivirus engines and machine learning detectors.

\subsection{Attacks on Android}
The increasing use of Android smartphones has led to a rise in the threat of Android malware, which makes cybersecurity a significant concern. It is crucial to adopt proactive detection and defense strategies to mitigate this issue. The paper~\cite{U5_rathore2023adversarial} is centered on developing adversary-aware Android malware detection models, addressing the escalating threat of malware on Android smartphones. It examines 36 models that were created using static features, such as permission and intent, along with different classification algorithms, with a focus on Android malware. The models for detecting malware are trained by utilizing static features, such as Android permission and intent, to take a proactive approach against the constantly evolving landscape of Android malware threats.

Wu et al.~\cite{U6_xu2023gendroid} addressed the need for a competent black-box adversarial attack structure for Android devices. The proposed solution, GenDroid, utilizes the SE-block attention model and JSMA algorithm to generate scalable adversarial examples, bypassing and misleading target detector classifiers. The authors introduced Gaussian process regression to optimize GenDroid, reducing queries to the target network and enhancing real-world applicability. Malicious code from Android source code: This proposed model utilized feature engineering, semantic enhancement, and classification techniques to detect malware~\cite{kar2022integrated}. Pretraining involves utilizing deep neural networks and large-scale data, and the output comments are merged with the decompiled source code vectors. This study demonstrated the effectiveness of the suggested hybrid models and their conceivable applications in the real world.

\subsection{Trespass on Internet of Things}

The challenge of detecting malware in IoT environments is a critical issue in the field of cybersecurity. The authors~\cite{U9_moti2021generative} proposed a novel approach using deep learning techniques, specifically GANs, to develop the accuracy and stability of classifiers for malware identification in IoT devices. After evaluating their models with standard IoT and Windows malware datasets, they reported promising results.

The problem stated in the paper~\cite{U11_ferrag2023generative} is the need for effective cybersecurity measures in the context of IoT networks (6G-enabled ), that are deliberately exposed to cyber attacks due to their complex and heterogeneous nature. A novel approach to cyber threat-hunting in networks is proposed, using a combination of a generative adversarial network and a Transformer-based model. This method harnesses the capabilities of generative AI to identify and avert attacks in real time.

Bakhsh et al.~\cite{U12_bakhsh2023enhancing} discussed the need for efficient Intrusion Detection Systems (IDS) to mitigate cyber threats in the rapidly growing IoT landscape. An IDS framework that utilizes deep learning techniques has been proposed to specify abnormalities in IoT networks. The framework includes a traffic capture mechanism that collects the traffic flow of IoT devices from the network. This collected data is then employed to simulate the processes of an IoT-enabled Smart Home. The proposed IDS system can detect and respond to cyber threats in IoT networks by identifying intrusions following sufficient training. There is a new algorithm for intrusion detection in edge networks using GANs. The algorithm is based on deep neural networks and aims to address security vulnerabilities within the Social Internet of Things (SIoT)~\cite{U13_de2020intrusion}. It tackles the ineffectiveness of traditional intrusion detection methods by employing GAN to generate realistic network traffic data for training a deep learning model. 

The study by Ferrag et al.~\cite{t9_ferrag2021federated} has emphasized the potential of security and privacy systems based on federated deep learning that can be used in different Internet of Things applications. These applications include Edge Computing, Cloud Computing, the Internet of Drones, Mobile Crowdsensing, the Industrial Internet of Things, 5G-enabled IoT, the Internet of Healthcare Things, and the Internet of Vehicles. Utilizing federated deep learning enhances the security and privacy of these networks by detecting cybersecurity attacks and mitigating the risk of data leakage. The incorporation of federated learning with blockchain technology further strengthens the security and control over stored and shared data in IoT networks. Hao et al.~\cite{hao2021asymmetric} concentrated on leveraging generative adversarial neural networks (GANs) for asymmetric cryptographic functions in the realm of the Internet of Things (IoT). Their proposed scheme bolsters IoT system security, addressing challenges like eavesdropping attacks, denial of service attacks, and database attacks through encryption and authentication across the control decision layer, smart communication layer, and perceptual excitation layer. The research conducted by Gupta et al.~\cite{t10_gupta2023chatgpt} explored core concepts of malware analysis, access control, AI, and machine learning-based cybersecurity and examined their practical implementation in various fields such as big data, cloud computing, cyber-physical systems, and the Internet of Things (IoT). The research aims to develop innovative security mechanisms, architectures, and models for smart cars, smart farming, and resourceful transportation policies, addressing potential vulnerabilities and designing robust security solutions against cyber threats in intelligent farming systems and critical agricultural infrastructure.

\subsection{Intrusion Detection System}
A comprehensive work~\cite{AS11_kumar2023synthetic}  addressed the challenges of data imbalance in intrusion detection systems (IDS). According to the authors, the proposed model can create robust attack signatures that can detect rare attacks even with limited publicly available samples. The paper highlights the model's versatility in various domains, such as medical diagnostics~\cite{zhuang2019lighter} and financial anomaly detection~\cite{baur2019cyber}, demonstrating its potential to improve detection performance using generative adversarial networks (GANs). Overall, the primary use case and application discussed in the paper involve using the model to develop an IDS that can effectively tackle data imbalance and identify novel attacks across domains.

Duy et al.~\cite{t1_duy2021digfupas} presented DIGFuPAS, a framework based on GANs that can generate adversarial samples against intrusion detection systems (IDS) in software-defined networking (SDN)-enabled networks. By producing attack traffic of various forms, this framework can assist in assessing and strengthening the resilience of IDS.

Shahriar et al.~\cite{t6_shahriar2020g} presented a GAN-supported IDS framework for cyber-physical systems (CPS) to enhance intrusion detection accuracy. Using GAN, the framework generates new data for labels requiring improved prediction. The findings and proposed framework have implications for strengthening intrusion detection and enhancing security in CPS, especially in emerging domains with limited data availability. An enhanced Network Intrusion Detection System (NIDS) that utilizes AI technology has been developed~\cite{t7_park2022enhanced}. This NIDS has successfully addressed the data imbalance issue and demonstrated improved classification performance compared to previous systems. Different data sets, consisting of standard data sets, an IoT data set, and an original data set, were used to evaluate the proposed system. The study demonstrated the feasibility of the proposed system in real-world environments and its potential application in detecting network threats efficiently.

\subsection{Face Recognition}

Face recognition technology is widely used in cybersecurity as a biometric authentication method. It provides enhanced security measures by uniquely identifying individuals based on their facial features. This technology finds applications in various areas, including access control, identity verification, and fraud prevention. The paper~\cite{U10_dirgantoro2020generative} discussed the limitations of a limited dataset in face recognition systems. The proposed system in the paper utilizes artificial intelligence and GANs)to improve the accuracy of the face recognition system. It is in contrast to traditional methods that rely solely on a limited dataset for training. Additionally, the proposed system utilizes edge computing to reduce latency and a low static difficulty blockchain to facilitate transactions.

\subsection{False Data Injection and Denial-of-Services}

Kar et al.~\cite{kar2022integrated} proposed an approach to detect and prevent cyber-attacks in the control of power systems. The researchers developed a deep neural network on GANs to identify and mitigate false data injection and denial-of-service attacks in power systems' broad-area blocking control loops. False data injection attacks are a type of cyber-attack where hackers can corrupt data streams from Phasor Measurement Units (PMUs) by gaining unauthorized access to the wide-area communication network channels. The researchers also addressed denial-of-service attacks, where attackers can temporarily deactivate a link to prevent PMU data from reaching its intended destination.

\subsection{Cyber Behavior Characterization}
A method that can be applied in a cyber behavior characterization~\cite{t8_ge2023explainable} system to describe cyber attack events and characterize attack behaviors. Analysts can identify the crucial attack techniques and comprehend the subsequent steps of the attack by breaking down CTI texts into individual entities and creating an attack behavior graph. It leads to the development of more sophisticated defense detection mechanisms, resulting in refined detection schemes.

\section{Security and Ethical Consideration}
Several research papers address critical security aspects in the field of cybersecurity. In~\cite{AS1_randhawa2021security}, the focus is on enhancing the security of botnet detection systems using Generative Adversarial Networks (GANs) to generate realistic botnet traffic samples. The aim is to improve detection performance and mitigate the effects of adversarial evasion attacks. This approach underscores the significance of using advanced techniques to address potential vulnerabilities in cybersecurity systems proactively. In~\cite{mazaed2022multifaceted}, researchers concentrate on improving the security of the Android operating system by designing a robust classification framework to protect public credentials and private data from malware attacks. Park et al.~\cite{t7_park2022enhanced} emphasize the need for enhanced network intrusion detection systems as cyber-attack techniques become increasingly complex and sophisticated. The expanding attack surface makes network systems more vulnerable to threats. Robust security measures are essential to counter these evolving challenges. Additionally, Kumar et al.~\cite{AS11_kumar2023synthetic} discuss the importance of rigorous testing and safeguarding sensitive data in the context of generating synthetic attack data for intrusion detection systems, highlighting the critical security implications of data generation in the cybersecurity domain.

Ethical considerations are integral to responsible and ethical research and application of ML, DL, and AI in cybersecurity. In the context of ChatGPT, as discussed in~\cite{t10_gupta2023chatgpt}, the focus is on addressing the vulnerabilities and limitations of the model. The study emphasizes that there is a requirement for enhanced security protocols to avoid unapproved entry and breach of data. This is in accordance with ethical standards of safeguarding user privacy and protecting data. It also highlights the importance of adhering to ethical guidelines and principles, such as the GDPR, in AI applications. In~\cite{t13_liu2023decompiling}, the extraction of DNN models from executables for adversarial example generation raises ethical concerns related to the potential misuse of models, emphasizing the importance of responsible model deployment. Similarly, in~\cite{AS8_won2022plausmal}, the potential for misuse of machine learning-based systems are mentioned, underscoring the significance of responsible usage and deployment protocols. The paper~\cite{AS9_reilly2023robustness}, while not explicitly mentioning ethical considerations, operates within the context of cybersecurity and machine learning, which inherently involve ethical dimensions related to privacy, data security, and potential biases in classification. Ethical diligence is crucial in the responsible application of AI technologies in the field of cybersecurity, striving for fairness, equity, and responsible data handling.

\section{Comparative Analysis}
\subsection{Methodology}
\subsubsection{GAN Methodology for Malware Detection}
\begin{figure}[ht]
\centering 
\includegraphics[width=.8\linewidth]{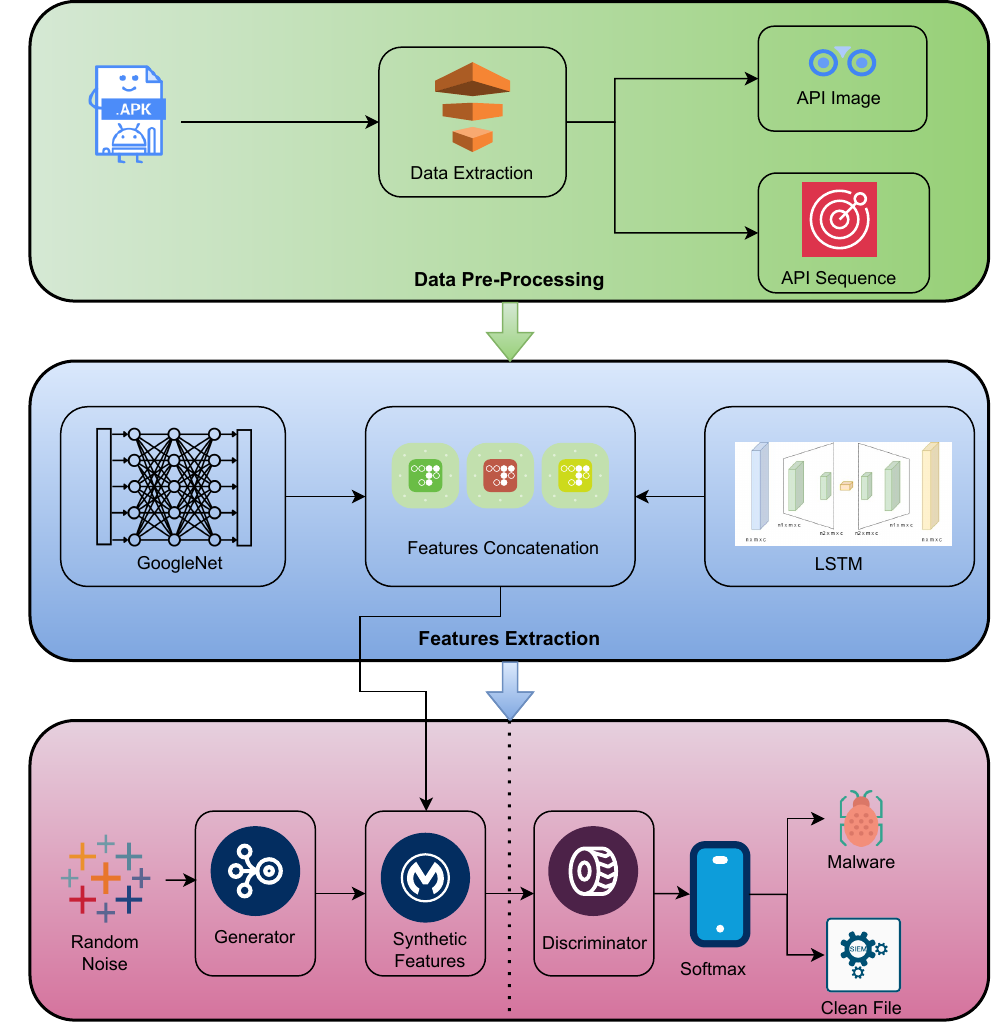}
\caption{The proposed framework for MDGAN~\cite{mazaed2022multifaceted}}
\label{fig:MD-gan} 
\end{figure}

Adversarial attacks and virus detection are two areas of cybersecurity where GANs have been widely used. GANs have demonstrated the potential to improve the precision and resilience of detection systems in the field of malware detection. Kim et al.~\cite{AS7_kim2022obfuscated} introduce an innovative approach to malware detection, combining global and local features to achieve an impressive 97.47\% accuracy. This exemplifies the potential of GANs in expanding the knowledge space and enhancing the precision of malware detection systems. Yang et al.~\cite{yang2022network} present an alternative method by merging Autoencoder with Adversarial Training, focusing on hiding malware and introducing unpredictability to thwart zero-day attacks. Mazaed et al.'s proposed Multifaceted Deep Generative Adversarial Networks Mode (MDGAN) architecture~\cite{mazaed2022multifaceted} combines Transformer and Conditional Generative Adversarial Network (CGAN) models for feature extraction and oversampling, showcasing the utility of GANs in addressing data imbalance. As shown in Figure~\ref{fig:MD-gan} we illustrated the proposed framework for the MDGAN.

Won et al.~\cite{AS8_won2022plausmal} delve into the application of GANs for creating malware images and enhancing classification models, highlighting the potential of GANs in bolstering security against sophisticated malware threats. Reilly et al.~\cite{AS9_reilly2023robustness} employ GANs to create adversarial instances for evaluating the resilience of malware detection models, exploring avenues to enhance robustness against evolving threats. Despite the divergence in methodologies, Ling et al.~\cite{U1_ling2023adversarial} address adversarial attacks against Windows PE malware detection systems, aligning with the overarching goal of testing and fortifying the resilience of malware detection systems. Lastly, Moti et al.~\cite{U9_moti2021generative} contrast their GAN-based malware detection method with alternative deep learning models, showcasing the efficacy of GANs in elevating accuracy, resilience, and adaptability to novel threats, thus significantly contributing to the advancement of malware detection.

\subsubsection{GAN Methodology for Intrusion Detections}
A nuanced exploration of GAN-based intrusion detection strategies reveals diverse approaches. Chui et al.~\cite{AS4_chui2023three} introduce the SMOTE-GAN-VAE methodology to address imbalanced classes in Network Intrusion Detection (NID), showcasing a novel integration of SMOTE, GAN, and VAE with specific CNN for feature extraction. Kumar et al.~\cite{AS11_kumar2023synthetic} leverage XGBoost-WCGAN for efficient minority attack data processing, outperforming traditional methods like Random Forest and Decision Tree. Furthermore, SVM. Bakhsh et al.'s DL-IDS framework employs deep learning approaches across various stages, enhancing Intrusion Detection System (IDS) efficiency in IoT networks. De et al.'s~\cite{U13_de2020intrusion} comparison highlights the superior accuracy and reduced false alarms of a GAN-based intrusion detection system, affirming its capacity to capture diverse attack types. Nie et al.~\cite{U14_nie2021intrusion} present the FID-GAN approach for Cyber-Physical Systems, achieving higher detection accuracy and lower false alarm rates. Collectively, these studies underscore the adaptability and efficacy of GAN-based methods in handling unbalanced data, detecting diverse attacks, and enhancing intrusion detection across varied scenarios.

\subsubsection{GAN Methodology for Adversarial Attacks}
Zhang et al.~\cite{AS3_zhang2020brute} introduce a brute-force attack method (BFAM), a new approach that generates adversarial examples in a black-box manner, showcasing alternative gradient-based methods~\cite{zhang2019extra, zhang2020novel, zhang2022extra}. Zhu et al.~\cite{U3_zhu2023adfl} utilize a GAN-based federated-learning approach to create synthetic samples for defending backdoor attacks. Lu et al.'s~\cite{U8_lu2019generative} integration of DCGAN with ResNet-18 for malware classification showcases GANs' versatility in synthetic data augmentation. These diverse approaches contribute to the broader landscape of cybersecurity research, addressing adversarial example generation, privacy concerns in federated learning, and synthetic data augmentation in malware classification.

\subsubsection{GAN Methodology for Cyber Threats}
The aforementioned research publications demonstrate how versatile GANs are in improving security measures by utilizing them in a variety of scenarios. With an emphasis on proactive evasion detection, GAN-based oversampling in Randhawa et al.~\cite{AS1_randhawa2021security} address data imbalance in botnet traffic detection, exemplifying proactive evasion detection. Soleymanzadeh et al.~\cite{soleymanzadeh2022stable} emphasize GANs' role in achieving high accuracy and stability in multi-classification models for intrusion detection. Dirgantoro et al.~\cite{U10_dirgantoro2020generative} showcase GANs' capability to overcome limited datasets by using dataset augmentation for improved face recognition accuracy. Yin et al.~\cite{yin2018enhancing} introduce Bot-GAN that increases labeled samples and highlights GANs' utility in enhancing detection capabilities. Ferrag et al.~\cite{U11_ferrag2023generative} accentuate the significance of GANs in creating synthetic data to boost model performance, particularly in detecting and preventing cyber-attacks in IoT networks. Collectively, these studies demonstrate the versatility and efficacy of GANs in addressing cybersecurity challenges, spanning cyber threat hunting, data augmentation, and intrusion detection in advanced IoT networks.

\subsection{Results and Findings}
In this section, significant findings are categorized by the cybersecurity domain. The major contributions to using Generative Adversarial Networks (GANs) for virus detection and classification are highlighted. Kim et al.~\cite{AS7_kim2022obfuscated} led a study that introduced a valuable method for detecting malware with an impressive accuracy of 97.47\%. This approach outperformed previous models and demonstrated superiority over many machine learning algorithms and deep neural networks. Won et al.~\cite{AS8_won2022plausmal} tested a novel GAN-based system for identifying zero-day malware, showcasing its effectiveness in distinguishing between known and zero-day threats even with limited training data. Mazaed et al.~\cite{mazaed2022multifaceted} proposed the MDGAN model as a superior method for detecting malware APIs on mobile devices, achieving higher precision, recall, F1-score, and average accuracy values compared to other methods. Visual-AT proved to be effective in malware detection, achieving high accuracy in cross-validation and outperforming recent tests~\cite{yang2022network}. Reilly et al.~\cite{AS9_reilly2023robustness} compared the byteplot and space-filling curve conversion methods for sorting malware images, revealing the vulnerability of the byteplot method to PGD attacks but improved resilience with DCGAN-generated adversarial images. Lu et al.~\cite{U8_lu2019generative} evaluated ResNet-18's ability to classify synthetic malware, while Babaagba et al.~\cite{babaagba2023evolutionary} illustrated malware's evasion of antivirus software, emphasizing the kNN model's superior performance. Kar et al.~\cite{kar2022integrated} presented a GAN-based approach for safeguarding power systems, demonstrating resistance to FDI and DoS attacks while maintaining stability during simultaneous attacks. Overall, these findings contribute to advancing methods for malware detection and classification, suggesting promising directions by leveraging GAN for future research and practical applications in cybersecurity.

Many studies' results and findings show that leveraging GAN has led to significant improvements in Intrusion detection systems. Network intrusion detection (NID) training data quality was enhanced by Chui et al.~\cite{AS4_chui2023three} using a three-stage data generation method with SMOTE, GAN, and VAE. The results showed significant improvements in accuracy on benchmark datasets. Kumar et al.~\cite{AS11_kumar2023synthetic} describe the XGBoost-WCGAN model, which performs better on the NSL-KDD, UNSW-NB15, and BoT-IoT datasets in terms of accuracy, recall, F1 score, and efficiency compared to other models. The statistical analysis proves that it has improved performance significantly, demonstrating its statistical importance. De et al.~\cite{U13_de2020intrusion} and Nie et al.~\cite{U14_nie2021intrusion} both wrote articles that compare GANs' success in cybersecurity to more traditional methods. The results show that GANs are more effective at identifying intrusion activities.DIGFuPAS, a GAN-based system for generating adversarial examples against IDS in SDN-enabled networks, is described by Duy et al.~\cite{t1_duy2021digfupas}. Their results show that IDS detection rates drop significantly when attacks are carefully planned. Shahriar et al.~\cite{t6_shahriar2020g} compare G-IDS (GAN-based IDS) to S-IDS and demonstrate that G-IDS is better at identifying attacks and maintaining stability. Das et al.~\cite{t11_das2022fgan} suggest a federated learning-based intrusion detection system that protects against model tampering, large-scale shared model changes, and data privacy by using encrypted communication. This makes intrusion detection systems more secure and private.

The authors of~\cite{U1_ling2023adversarial} provide an assessment of attack performance, outperforming previous attacks in terms of query efficiency and evasion rate. Proposed defenses are presented, offering a thorough analysis and suggestions for further research. These include a feature selection technique and ensemble learning. Furthermore,~\cite{U3_zhu2023adfl} demonstrates the effectiveness of ADFL in protecting federated learning against backdoor attacks, attaining an astounding 95\% decrease in attack success rates with primary task accuracy continuously above 90\%. According to the study in~\cite{t4_schneider2023dual}, there is a lot of misunderstanding and doubt when it comes to users' capacity to recognize hostile samples. Moreover,~\cite{t12_lucas2023adversarial} explores adversarial training for raw-binary malware classifiers and shows how various techniques might lower attack success rates. Concurrent training on many adversarial examples improves the overall robustness of the model.

In the area of cyber threat detection, Ferrag et al.~\cite{U11_ferrag2023generative} offer a GAN and Transformer-based model specially made for 6G-enabled IoT networks. With an overall accuracy of 0.95, the model successfully detects and classifies cyber threats with a noteworthy 95\% testing accuracy, demonstrating steady performance improvement over time.  In a different study, Ferrag~\cite{t9_ferrag2021federated} examines a variety of datasets and shows that neural network models, particularly LSTM, perform better in IoT security than simple machine learning models. The models perform well in all classes, but they stand out in DDoS and CnC situations the most. Furthermore, Ferrag~\cite{t9_ferrag2021federated} uses Speech Command and ECG datasets on Raspberry Pi to explore federated deep learning for IoT security. The research indicates that after 50 rounds, the global model performance has increased, confirming the effectiveness of federated deep learning in augmenting IoT security. 
Dirgantoro et al.~\cite{U10_dirgantoro2020generative} offer valuable perspectives on facial recognition precision through GAN-generated data. 
They report enhanced accuracy and present a low static difficulty blockchain that demonstrates superior transaction time compared to Proof-of-Work. Bakhsh et al.~\cite{U12_bakhsh2023enhancing} compare FFNN, LSTM, and RandNN with machine learning techniques in order to evaluate the generalization ability of deep learning models in threat detection. Gupta et al.~\cite{t10_gupta2023chatgpt} contrast ChatGPT with Google Bard's cybersecurity features and highlight the latter's potential for threat detection. While Google Bard performs exceptionally well at identifying Union SQL injections and providing prompt remedial actions, ChatGPT is superior at detecting other types of SQL injections. For CTI text analysis, Ge et al.~\cite{t3_ge2023explainable} present a self-adversarial topic generation model that outperforms previous models in evidence boundaries, semantic evidence, and stability while also earning higher F1 scores. Last but not least, Hao et al.~\cite{hao2021asymmetric} investigate cyber-physical network security through double-DES encryption, proving the effectiveness of the plan against eavesdropping and man-in-the-middle attacks. Randhawa et al.~\cite{t5_randhawa2022evasion} demonstrate the superiority of EVAGAN over ACGAN in detecting anomalies in computer networks and improve AI-based anomaly detection through effective model training with fewer samples.

We provide a comprehensive overview of study results from several cybersecurity fields in Table~\ref{tab:findings}. Within the realm of botnets, techniques such as GAN-based Oversampling and the Bot-GAN Framework demonstrate increased recall scores while demonstrating better accuracy and greater botnet detection, respectively. GAN-based frameworks and models for malware detection show excellent accuracy in identifying zero-day threats and virtual malware. Evaluations of adversarial attacks point to better query efficiency and evasion rates. GAN-based and transformer-based models, and other neural network-based models, have shown efficacy in detecting and categorizing cyber risks in the area of the Internet of Things. Other discoveries include increased facial recognition accuracy with data generated by GANs, the ability of deep learning models to generalize in threat detection, and intriguing capabilities in cybersecurity tools such as ChatGPT and Google Bard. The table summarizes a variety of research projects and offers a succinct summary of the most important cybersecurity research findings.

\begin{table*}[th]
    \centering
    \caption{This table captures GAN-based cybersecurity research findings. From enhanced botnet detection using GANs to improved malware detection frameworks and advancements in adversarial attack defenses, it provides a comprehensive overview. Additionally, insights into IoT security and the effectiveness of models like ChatGPT in cybersecurity underscore the evolving landscape of cyber threats and defense strategies.}
    \begin{tabular}{|l|l|p{5cm}|l|}
    \hline
    \textbf{Domain} & \textbf{Method/Model} & \textbf{Findings} & \textbf{Reference} \\
    \hline
    Botnet and IDS & GAN-based Oversampling & Improved accuracy; impact on recall scores, effective in evading classifiers. &~\cite{AS1_randhawa2021security} \\
    \cline{2-4}
    & Bot-GAN Framework & Enhanced botnet detection, synthetic samples, improved metrics. &~\cite{yin2018enhancing} \\
    \cline{2-4}
    & Three-stage Data Generation Algorithm & Improved training data quality, notable accuracy enhancements. &~\cite{AS4_chui2023three} \\
    \cline{2-4}
    & XGBoost-WCGAN Model & Outperformed baselines, superior metrics. &~\cite{AS11_kumar2023synthetic} \\
    \cline{2-4}
    & DIGFuPAS Framework & Significant decrease in IDS detection rates, preserving network features. &~\cite{t1_duy2021digfupas} \\
    \cline{2-4}
    & G-IDS vs. S-IDS & G-IDS exhibited superior attack detection, stable F1 scores. &~\cite{t6_shahriar2020g} \\
    \cline{2-4}
    & Federated Learning-based IDS & Enhanced security, privacy in intrusion detection systems. &~\cite{t11_das2022fgan} \\
    \hline
    Malware & Virtual Malware Detection & High accuracy, outperforming traditional models. &~\cite{AS7_kim2022obfuscated} \\
    \cline{2-4}
    & GAN-based Framework for Zero-Day & Exceptional accuracy in detecting zero-day malware. &~\cite{AS8_won2022plausmal} \\
    \cline{2-4}
    & MDGAN for Malware APIs & Demonstrated high effectiveness in detecting malware APIs. &~\cite{mazaed2022multifaceted} \\
    \cline{2-4}
    & MalGAN for Malware Detection & Outperformed other methods in accuracy. &~\cite{U9_moti2021generative} \\
    \cline{2-4}
    & Robustness Evaluation of Classification & Improved robustness with adversarial images. &~\cite{AS9_reilly2023robustness} \\
    \hline
    Adversarial Attack & Adversarial Attack Evaluation & Outperformed existing attacks in evasion rate and query efficiency. &~\cite{U1_ling2023adversarial} \\
    \cline{2-4}
    & ADFL in Federated Learning & Significant reduction in attack success rates maintained high accuracy. &~\cite{U3_zhu2023adfl} \\
    \cline{2-4}
    & Dual Research Questions on AS & Users exhibited confusion in discerning adversarial samples. &~\cite{t4_schneider2023dual} \\
    \cline{2-4}
    & Adversarial Training for Raw-Binary & Enhanced overall model robustness with simultaneous training on adversarial examples. &~\cite{t12_lucas2023adversarial} \\
    \hline
    IoT & GAN and Transformer-based Model & Achieved high testing accuracy, accurately identified and classified cyber threats. &~\cite{U11_ferrag2023generative} \\
    \cline{2-4}
    & Federated Deep Learning for IoT Security & Improved global model performance, effective in IoT security. &~\cite{t9_ferrag2021federated} \\
    \hline
    Miscellaneous & Face Recognition with GANs-generated Data & Improved face recognition accuracy with GANs-generated data. &~\cite{U10_dirgantoro2020generative} \\
    \cline{2-4}
    & Generalization Capacity of DL Models & Evaluated generalization capacity of deep learning models in threat detection. &~\cite{U12_bakhsh2023enhancing} \\
    \cline{2-4}
    & Comparison of ChatGPT and Google Bard & Both showed promise in detecting and addressing cyber threats. &~\cite{t10_gupta2023chatgpt} \\
    \cline{2-4}
    & Self-Adversarial Topic Generation Model & Higher F1 scores, the outperformance of other models in evidence boundary and stability. &~\cite{t3_ge2023explainable} \\
    \cline{2-4}
    & Cyber-Physical Network Security & Demonstrated resistance to attacks, ensuring security in cyber-physical networks. &~\cite{hao2021asymmetric} \\
    \cline{2-4}
    & GAN Models in Anomaly Detection & EVAGAN demonstrated superior stability and qualitative performance over ACGAN. &~\cite{t5_randhawa2022evasion} \\
    \hline
    \end{tabular}
    \label{tab:findings}
\end{table*}

\subsection{Datasets}
This section offers a comprehensive overview of the various datasets that are frequently used in GAN-based cybersecurity research. These datasets are essential for assessing and developing cybersecurity solutions, covering tasks like malware analysis, intrusion detection, Internet of Things (IoT) security, and mitigating adversarial attacks. The distinct features and obstacles presented by every dataset aid in the improvement and verification of machine learning models intended to counteract cyberattacks.

Researchers have been actively investigating the possibility of using Generative Adversarial Networks (GANs) to strengthen botnet detector defenses in the ever-changing field of cybersecurity. An extensive analysis of this field was conducted by~\cite{AS1_randhawa2021security}, with particular attention to leveraging the ISCX-2014 dataset. Using GANs to improve the dataset, the researchers thoroughly assessed the six classifiers' performance in their study. In addition, a Blackbox test was run to determine how well the generated botnet data eluded detection systems.

A complex three-stage data creation approach was presented in~\cite{AS4_chui2023three}, which addresses the problem of imbalanced datasets in network intrusion detection (NID). This novel method made use of the capabilities of the Variational Autoencoder (VAE), Generative Adversarial Network (GAN), and Synthetic Minority Oversampling Technique (SMOTE). The effectiveness of the approach was evaluated on four benchmark datasets: NSL-KDD, UNSW-NB15, KDD Cup 1999, and CICIDS2017. Notable is the nuanced assessment, which introduced another level of complexity to the accuracy comparison by taking into account the regrouping of attack types in CICIDS2017.

Kumar et al.~\cite{AS11_kumar2023synthetic} introduce an enhanced model, XGBoost-WCGAN, in intrusion detection tasks, intending to produce data for training intrusion detection systems. The model's performance was carefully assessed on three different datasets: NSL-KDD, UNSW-NB15, and BoT-IoT. The study demonstrated the model's flexibility and efficacy in various cybersecurity scenarios.

A thorough investigation was carried out by~\cite{U8_lu2019generative} into the constantly changing field of malware detection. With the help of artificial malware samples produced by the Deep Convolutional Generative Adversarial Network (DCGAN), the study assessed the classification accuracy of a deep residual network (ResNet-18). The Malimg dataset, a collection of 9,391 malware samples from 25 distinct families, was used for the assessments. This thorough analysis shed light on the model's capacity to identify and categorize various malware kinds.

There is an increasing requirement for strong defenses against hostile attacks as the danger landscape grows. Contributions to this field were made by~\cite{U3_zhu2023adfl}, who carried out experiments on a dataset of 70,000 handwritten images from the MNIST dataset, each normalized to 28x28 pixels. The study shed important light on how well the suggested defense systems fend off hostile attacks.

\begin{table*}[H]
\centering
\caption{Overview of diverse datasets extensively used for GAN in cybersecurity research. Ranging from intrusion detection tasks to malware analysis and adversarial attacks, these datasets serve as fundamental resources for evaluating and advancing cybersecurity solutions.}
\label{Datasets}
\resizebox{\textwidth}{!}{%
\begin{tabular}{|p{3.5cm}|p{11.5cm}|}
  \hline        
  \textbf{Data Set Name} & \textbf{Data Set Description} \\
  \hline  
  KDD Cup 1999~\cite{AS4_chui2023three} & This dataset was used for the intrusion detection task in the Third International Knowledge Discovery and Data Mining Tools Competition (KDD Cup 1999). The dataset from the KDD Cup 1999 has about 4.9 million instances.\\
  \hline  
  NSL-KDD~\cite{AS4_chui2023three, AS11_kumar2023synthetic} & The KDD Cup 1999 dataset was enhanced and is now known as the NSL-KDD dataset. There are 125,973 instances in total. The purpose of this dataset is to assess computer network intrusion detection systems. It includes DoS, Probe, R2L (Unauthorized access from a remote machine), and U2R (Unauthorized access to local superuser privileges) among other types of attacks.\\
  \hline  
  UNSW-NB15~\cite{AS4_chui2023three, AS11_kumar2023synthetic} & About 2.5 million instances, including both legitimate and malicious network traffic, make up the UNSW-NB15 dataset. This is a dataset for network intrusion detection that covers a broad spectrum of attacks, including denial-of-service, analysis, exploitation, and reconnaissance. \\
  \hline          
  BoT-IoT~\cite{AS11_kumar2023synthetic} & A benchmark dataset for assessing the security of IoT devices is called BoT-IoT. It provides realistic network traffic patterns and security vulnerabilities for more than 20 different types of devices. The dataset is essential for testing anomaly detection algorithms and intrusion detection systems, which advance IoT security solutions.\\
  \hline          
  MNIST dataset~\cite{U3_zhu2023adfl} & MNIST is a fundamental dataset for image classification, containing 70,000 grayscale images of handwritten digits (0-9). With 28x28 pixel resolution, it serves as a benchmark for assessing the performance and generalization of machine learning models. MNIST's simplicity and standardization make it a popular choice for testing and comparing image recognition algorithms.\\
  \hline          
  Malimg dataset~\cite{U8_lu2019generative}& The Malimg dataset is a collection of grayscale images representing various families of malware. It serves as a resource for researching malware detection through image-based analysis. Researchers use this dataset to explore visual patterns and characteristics associated with different types of malware, enabling the development and evaluation of image-based classification algorithms.\\
  \hline 
  CV MNIST~\cite{t5_randhawa2022evasion} & The CV MNIST dataset, which consists of 28x28 pixel grayscale pictures of handwritten numbers 0 through 9, is a benchmark in computer vision. 70,000 photos in total, divided into testing and training sets. Since each image has the matching digit labeled on it, it's a perfect option to evaluate how well different image recognition models perform.\\
  \hline 
  CIC-DDoS2019~\cite{U13_de2020intrusion} & The CIC-DDoS2019 dataset is designed for researching Distributed Denial of Service (DDoS) attacks. It offers a diverse collection of network traffic data under both normal and DDoS attack conditions. With a significant number of samples, this dataset provides valuable insights into DDoS attack characteristics, including flow statistics, protocol distribution, and payload details. \\
  \hline           
  CSE-CIC-IDS2018~\cite{U13_de2020intrusion} & For IDS research, a sizable cybersecurity dataset called CSE-CIC-IDS2018 was created. It is flexible for assessing intrusion detection systems since it includes more than 16 million records containing a combination of legitimate and malicious activity. The dataset includes important details like IP addresses, port numbers, protocols, and timestamps and covers a range of cyber threats, including DDoS and DoS attacks. \\ \hline
  IoT-23~\cite{garcia2020iot23} & The purpose of the IoT-23 dataset is to evaluate intrusion detection systems within the Internet of Things (IoT). It includes twenty-three different cyberattack scenarios, providing approximately 12,000 examples in total. This dataset is a useful tool for improving cybersecurity in IoT environments because it makes it possible to evaluate detection algorithms across a range of IoT device types, communication protocols, and malicious activities.\\
  \hline            
  CC botnet~\cite{t5_randhawa2022evasion} & A collection of network traffic data called the CC botnet dataset is used to research and examine the communication patterns of botnets. This dataset, which consists of labeled examples of botnet traffic, is an important tool for researchers who want to improve and build models for the identification of malicious botnet activity. \\ \hline 
\end{tabular}%
}
\end{table*}

Overall, the descriptions of the datasets in Table~\ref{Datasets} provide a thorough summary of a variety of datasets that are essential to the development of GAN for machine learning and cybersecurity models. Notably, with its large number of examples, the KDD Cup 1999 dataset—which was created for intrusion detection—lays the groundwork for assessing security measures. Building on this, the NSL-KDD dataset used in the experiment to improve intrusion detection assessments by emphasizing various attack kinds. The UNSW-NB15 dataset makes a substantial contribution to intrusion detection research since it includes a wide range of network traffic. BoT-IoT is a benchmark dataset that makes it easier to assess IoT device security solutions by using actual traffic patterns. A key reference point for testing and evaluating machine learning models is the MNIST dataset, which is essential to image categorization. The Malimg dataset fosters the development of classification algorithms by offering a distinct viewpoint on malware identification through image-based research. The CIC-DDoS2019, CSE-CIC-IDS2018, and CV MNIST datasets provide detailed insights into intrusion detection, DDoS attacks, and computer vision, respectively. Finally, the IoT-23 dataset adds scenarios covering a range of device kinds and communication protocols to the cybersecurity landscape. It is specifically designed for IoT intrusion detection. Every dataset is essential to the advancement of innovation and study in the constantly changing realm of cybersecurity.

\subsection{Evaluation Metrics}
In the field of cybersecurity research, a variety of assessment measures are essential for determining how effective certain strategies are. The particular task at hand determines which metrics are used, and each statistic has a distinct function in measuring the effectiveness of the suggested approaches.

Recall is a common evaluation metric. Recall is highlighted as a crucial statistic in the study by Randhawa et al.~\cite{AS1_randhawa2021security}, highlighting its significance in reducing the consequences of hostile evasions. Notably, it is still difficult to achieve meaningful recall improvement for all classifiers on various datasets. The equation for calculating recall is 

\begin{equation}
\text{Recall} = \frac{\text{True Positive (TP)}}{\text{True Positive (TP) + False Negative (FN)}}
\end{equation}

According to several research~\cite{AS1_randhawa2021security, AS7_kim2022obfuscated, AS11_kumar2023synthetic}, precision is another commonly used metric. Precision is a useful metric for activities where false positives might have substantial repercussions, as it gauges the accuracy of positive predictions. Precision, for instance, is a crucial parameter for evaluating the suggested model's capacity to correctly identify dangerous occurrences in malware detection~\cite{AS7_kim2022obfuscated}. Precision can be defined as follows: 

\begin{equation}
\text{Precision} = \frac{\text{True Positive (TP)}}{\text{True Positive (TP) + False Positive (FP)}}
\end{equation}

Many cybersecurity studies use the accuracy evaluation metric~\cite{AS4_chui2023three, AS7_kim2022obfuscated, t2_liu2020efficient}. Because accuracy measures the entire correctness of predictions, it offers a comprehensive picture of model performance. It is particularly pertinent to activities like intrusion detection, where it is essential to analyze both benign and malevolent cases accurately.

\begin{equation}
\text{Accuracy} = \frac{\text{Number of Correct Predictions}}{\text{Total Number of Predictions}}
\end{equation}

Studies in the field of malware detection frequently use a variety of measures, such as p-value, accuracy, F1-score, and standard deviation~\cite{AS7_kim2022obfuscated}. The model's performance across various malware kinds is comprehensively evaluated by the F1-score, which strikes a balance between precision and recall.

\begin{equation}
p \text{-value} = P(\text{observing test statistic} \geq \text{observed value} | H_0)
\end{equation}

Several metrics are relevant for intrusion detection jobs. The most often used metrics are recall, FAR, precision, and F1 score~\cite{AS11_kumar2023synthetic, U13_de2020intrusion, t6_shahriar2020g}. Together, these measures evaluate how well the model can identify and categorize intrusions while reducing false alarms.

\begin{equation}
     F1 \text{-} \text{score} = \frac{2 \times (\text{Precision} \times \text{Recall})}{\text{Precision} + \text{Recall}} 
\end{equation}

\begin{equation}
\text{FAR} = \frac{\text{False Positives}}{\text{False Positives + True Negatives}}
\end{equation}

In order to evaluate the efficacy of the created malware samples, Won et al.'s paper~\cite{AS8_won2022plausmal} presents two novel evaluation metrics: the Quality Fitness Score and the Diversity Fitness Score. By taking into account both the quality and diversity of the generated samples, these metrics offer a sophisticated assessment.

Evaluation metrics such as Evasion Increase Rate (EIR) and Detection Rate (DR) are important in the context of adversarial attacks~\cite{t1_duy2021digfupas}. The black-box IDS's detection capability is gauged by DR, while EIR evaluates the framework's capacity to produce adversarial samples that are undetectable.

\begin{equation}
    EIR = \frac{{\text{{Number of misclassified adversarial samples}}}}{{\text{{Total number of adversarial samples}}}}
\end{equation}

\begin{equation}
\text{DR} = \frac{\text{Number of correctly detected attack traffic}}{\text{Total amount of incoming attack traffic}}
\end{equation}

In the context of GANs, another evaluation metric that is frequently used to evaluate image generation quality and variety is the Fréchet Inception Distance (FID)~\cite{AS8_won2022plausmal}. Feature extraction from an Inception-v3 deep neural network is used to quantify the similarity between the distribution of generated and real images. Stated differently, the Fréchet Inception Distance (FID) is a statistical measure of how similar generated images are to real images based on distributional and statistical properties. Higher quality and greater diversity in the generated images are indicated by a lower FID score. The FID metric has been widely adopted in the fields of image generation and Generative Adversarial Network (GAN) assessment. The FID equation is 

\begin{equation}
    \text{FID} = \|\mu_r - \mu_g\|^2 + \text{Tr}(\Sigma_r + \Sigma_g - 2(\Sigma_r\Sigma_g)^{1/2})
\end{equation}

Here, \(\mu_r\) and \(\mu_g\) are the mean vectors of the features from real and generated samples, respectively. \(\Sigma_r\) and \(\Sigma_g\) are the covariance matrices of the features from real and generated samples, respectively. \(\text{Tr}(\cdot)\) denotes the trace of a matrix. \(\|\cdot\|\) represents the Euclidean distance.

Numerous metrics, including Detection Rate (DR), Structural Similarity (SS), Behavioral Similarity (BS), ROC AUC Score, and more, are used in other research~\cite{babaagba2023evolutionary, t5_randhawa2022evasion, t9_ferrag2021federated}. These metrics provide a thorough grasp of the models' performance by addressing particular subtleties of the jobs.

\begin{table*}[!t]
\centering
\caption{Evaluation Metrics}
\label{table:gan-metric}
\renewcommand{\arraystretch}{1.5} 
\begin{tabular}{p{5cm}p{8cm}}
\hline
\textbf{GAN} & \textbf{Evaluation Metrics}\\  
\cline{1-2}
Vanilla \& CGAN~\cite{AS1_randhawa2021security} & Cross Validation, Recall, Accuracy, Precision, F1 Score\\ 
WGAN \& SMOTE-GAN-VAE~\cite{AS4_chui2023three} & Accuracy, Cross Validation\\ 
WGAN~\cite{AS7_kim2022obfuscated} & Accuracy, P-Value, F1-Score\\ 
PlausMal-GAN~\cite{AS8_won2022plausmal} & Std. dev., Acc, Frechet inception distance (FID)\\ 
WCGAN~\cite{AS11_kumar2023synthetic} & Precision, Recall, F1 Score, FAR (False Alarm Rate)\\ 
MDGAN~\cite{mazaed2022multifaceted} & Mean Precision, Mean Recall, Mean F-Score\\  
\hline
\end{tabular}
\end{table*}

To summarize, the selection of assessment metrics in cybersecurity research is contingent upon the specific task at hand, with crucial roles being played by precision, recall, accuracy, F1-score, and task-specific metrics. The variety of indicators is a reflection of the complexity of cybersecurity issues and the demand for detailed performance evaluations. Table~\ref{table:gan-metric} provides a concise overview of various Generative Adversarial Networks (GANs) used in cybersecurity research along with the associated evaluation metrics. 

\subsection{Limitation}
Navigating the multifaceted landscape of Generative Adversarial Networks (GANs) within the realm of cybersecurity necessitates a thorough exploration of challenges and limitations. This segment provides insights into issues related to adversarial attacks, data imbalance, and the intricacies associated with creating authentic cyber threats. Additionally, it critically evaluates limitations intrinsic to GAN variation, training instability, and broader perspectives within the cybersecurity domain.

In the study conducted by Radhawa et al.~\cite{AS1_randhawa2021security}, both Vanilla GAN and Conditional GAN were employed for the generation of botnet traffic. Vanilla GAN's limitation arises from its reliance on noise input during sample creation, potentially resulting in outputs of lower quality or lacking specific features. To address this, Conditional GAN enhances control by incorporating additional characteristics during the generation process. However, the effectiveness of Conditional GAN hinges on the precision of extra labels, introducing challenges if these inputs fail to capture the nuances of actual data accurately. Wasserstein GAN (WGAN), renowned for generating adversarial instances in cybersecurity~\cite{AS3_zhang2020brute}, grapples with training instability. The quality and diversity of training data significantly impact this instability, making it challenging to generate reliable adversarial examples~\cite{t20_lee2021gan}. The black-box nature of WGAN poses challenges in terms of adversary intervention in the generation process, complicating targeted perturbations required in certain cybersecurity scenarios.

In the realm of interpretable representation learning, InfoGAN~\cite{t8_ge2023explainable} assumes the independence and Gaussian distribution of latent variables. While mathematically tractable, this assumption may deviate from actual data distributions, affecting InfoGAN's performance. Moreover, InfoGAN requires substantial training data, and its efficacy diminishes when confronted with smaller datasets.

The utilization of FedGAN~\cite{t9_ferrag2021federated} in federated deep learning for IoT networks introduces potential risks, with adversarial attacks and data poisoning threatening model accuracy. These challenges underscore the importance of robust security measures. Dataset limitations, as evidenced in~\cite{AS4_chui2023three}, emphasize the need for improved training data quality to address issues in Network Intrusion Detection (NID) models. Although evaluations are constrained to specific datasets, novel approaches, such as GenDroid~\cite{U6_xu2023gendroid}, exhibit promise in Android malware detection. Exploring the GenDroid architecture with more sophisticated models and features could further enhance its capabilities.

Several research works~\cite{t17_yilmaz2020addressing}~\cite{t18_liu2021gan},~\cite{AS11_kumar2023synthetic}~\cite{t19_nazari2021using} draw attention to challenges related to class imbalance, dataset representativeness, and the need for comprehensive applicability evaluations. Complete transparency regarding data privacy protections and ethical considerations is imperative. The study by~\cite{AS9_reilly2023robustness} highlights computational resource challenges, including limitations on GPU availability and session durations. These constraints impact the scope and depth of experimental efforts, emphasizing the persistent challenges associated with computational requirements.

Overall, this survey presents many unique sets of limitations, encompassing evaluation techniques, dataset constraints, and considerations of real-world applicability. These drawbacks underscore the dynamic nature of GAN applications in cybersecurity, emphasizing the ongoing need for investigation and improvement in these cutting-edge models.

\section{Future Directions}

Generative adversarial networks are responsible for the generation of synthetic data when they incorporate intrinsic variability. This has an effect on the consistency and quality of the samples that are created. As per the findings of the research conducted by Radhawa and colleagues~\cite{AS1_randhawa2021security}, the presence of this variability presents difficulties when it comes to the generation of genuine hostile botnet traffic for the purpose of training intrusion detection systems. The development of specialized GANs that are specifically created for the purpose of creating adversarial botnet traffic should be the focus of future research. The goal of this research should be to imitate the behavior of actual botnet environments closely. Intrusion detection systems are able to improve their accuracy against sophisticated botnet operations if they are trained with data that is truly representative of the current environment.

According to Zhang et al.~\cite{AS3_zhang2020brute}, another problem is the potential instability of GANs during training, which can have an impact on the creation of samples that are both dependable and diverse. The effectiveness of intrusion detection models may be negatively impacted as a result of this instability, particularly when it comes to discriminating between botnet traffic and regular network traffic. To overcome this issue, future research should make use of GANs to increase feature development, which will, in turn, enhance the ability of intrusion detection systems to differentiate between different sources of intrusion. Furthermore, continued efforts to investigate sophisticated adversarial examples created by GANs might function as a constant challenge, which in turn can drive advancements in the robustness of these systems.

It is possible that GANs may encounter constraints when it comes to the generation of different characteristics in the field of malware detection, which might potentially impede the performance of detection systems~\cite{AS8_won2022plausmal} for example. The proposed strategy for the future entails making use of GANs in order to develop synthetic malware samples that cover a variety of different categories. These varied datasets prove to be extremely useful when it comes to the training and evaluation of malware detection algorithms. Furthermore, GANs have the potential to play a crucial role in the production of sophisticated features. This is accomplished by integrating visual representations and data sequence features in order to enhance the classification performance of malware detection frameworks~\cite{mazaed2022multifaceted}.

\begin{table*}[h]
\caption{Future Directions for Addressing Limitations}
\label{tab:future-recommendation}
\begin{tabularx}{\textwidth}{|l|X|}
\hline
Limitation                                                              & Future Work/Recommendation                                                                                                                                                                                                                                               \\ \hline
Inherent Variance                                                       & Adversarial Botnet Generation: Create GANs with a focus on generating adversarial botnet traffic to improve training dataset realism and tackle GAN variance problems.                                                                                                   \\ \hline
\multirow{3}{*}{Training Instability in Botnet and Intrusion Detection} & Better Feature Generation: Use GANs to provide useful features that help identify botnet traffic and improve intrusion detection models' discriminative power.                                                                                                           \\ \cline{2-2} 
                                                                        & Enhanced Adversarial samples: Investigate GAN-generated enhanced adversarial instances to continuously test and enhance intrusion detection system resilience while reducing  training instability.                                                                      \\ \cline{2-2} 
                                                                        & Autonomous GAN-Based IDS: Use deep reinforcement learning to integrate GANs into intrusion detection systems for autonomous model development and refinement, with a special emphasis on adjusting to new evasion examples. This may improve the stability of the model. \\ \hline
\multirow{2}{*}{Limited Feature Extraction in Malware Detection}        & Generating Malware using GANs: GANs can be used to generate synthetic malware samples of different kinds, offering a varied and difficult dataset for testing and training malware detection systems.                                                                    \\ \cline{2-2} 
                                                                        & Advanced Feature Generation: Use GANs to increase feature extraction from malware samples. To boost classification performance, combine data sequence features with visual representations. This deals with the difficulties in feature extraction.                      \\ \hline
\multirow{2}{*}{Robustness Against Adversarial Attacks}                 & Adversarial Example Generation: Create GANs with a focus on producing sophisticated adversarial instances in order to test the robustness of various machine learning classifiers.                                                                                       \\ \cline{2-2} 
                                                                        & GAN-Based Attack Comparative Analysis: Examine the efficacy of GAN-based attack techniques and contrast them with other GAN-based techniques in a range of cybersecurity contexts. This strengthens defense systems against hostile attacks.                             \\ \hline
Data imbalance                                                          & Using GANs to Mitigate Data Imbalance: To solve data imbalance in cybersecurity datasets and improve the efficacy of machine learning models, use GANs to generate synthetic samples.                                                                                    \\ \hline
\multirow{3}{*}{Model Performance and Adaptability}                     & GAN-Based Transfer Learning: To improve model performance and flexibility, look into using GANs in transfer learning for network intrusion detection.                                                                                                                    \\ \cline{2-2} 
                                                                        & GAN-Based Model Refinement: By integrating GANs into intrusion detection systems, feature extraction and model refinement may be optimized, increasing accuracy and resilience to new threats.                                                                           \\ \cline{2-2} 
                                                                        & Adversarial Defense using GANs: To ensure security against different evasion techniques, use GANs to develop intrusion detection models resistant to adversarial attacks.                                                                                                \\ \hline
\end{tabularx}
\end{table*}

There is a possibility that machine learning classifiers, which encompass intrusion detection systems, could potentially be susceptible to sophisticated adversarial attacks~\cite{AS3_zhang2020brute}. The development of specialized GANs for the generation of sophisticated adversarial cases should be the primary focus of future research initiatives. This will contribute to the evaluation and improvement of classifier robustness. It is possible to gain more insights into efficient security mechanisms by doing a comparison analysis of GAN-based attack strategies. In addition, it is advised that GANs be used to address data imbalance as a potential future option. This would ensure classifiers handle imbalanced datasets more effectively.

Based on the findings of the research conducted by Chui et al.~\cite{AS4_chui2023three}, it is possible that intrusion detection models may encounter difficulties in terms of their overall performance and adaptability. This is especially true when they are confronted with new threats and different datasets. Researchers are strongly urged to investigate the possibility of incorporating GANs into intrusion detection systems to improve model accuracy and provide adversarial defense. By taking this method, the goal is to improve protection against changing threats while also optimizing the extraction of features. It will be possible to improve the application of GANs in cybersecurity by resolving these constraints through the next studies that are advised. This will result in security systems that are more efficient and resistant to attacks.
A comprehensive future direction for the above limitations has been portrayed in Table~\ref{tab:future-recommendation}.

\section{Conclusion}
Finally, the important role that Generative Adversarial Networks (GANs) play in enhancing cybersecurity defenses has been examined in this survey research. It is now essential to have cybersecurity-based solid defense measures in place due to the growing volume of data that must be collected and sent digitally. GANs in particular, which are deep learning models, have shown to be effective tools for tackling the always-changing security issues in the digital world.

In this analysis, we have looked at a variety of GAN applications in cybersecurity areas, including malware detection, mobile and network trespass, intrusion detection systems (IDS), and botnet detection. The results have demonstrated how GANs may strengthen cybersecurity defenses in these domains. By producing realistic cyber data for machine learning classifier training and evaluation, GANs have demonstrated the potential to raise intrusion detection systems' efficacy and accuracy. They have also proven successful in identifying malware and spotting irregularities in network traffic.

It's crucial to recognize the difficulties and limitations that come with using GANs in cybersecurity, though. Some of the major issues that need to be resolved are evasion strategies, data accuracy, imbalanced data, computing cost, zero-day malware, and adversarial attacks. Subsequent investigations ought to concentrate on enhancing GAN models' resilience against hostile instances, tackling resource limitations, and refining methods for preparing data.

In addition, ethical and responsible data handling is essential when using GANs in cybersecurity. To stop the spread of propaganda and false information, it is crucial to ensure the appropriate and responsible usage of GANs, as they have the potential to produce fake media and mislead both human and machine judgment.

In conclusion, this survey has offered a thorough examination of the present status of research on the use of GANs in cybersecurity. It has outlined the difficulties and limitations that must be overcome while emphasizing the potential advantages of GANs in bolstering cybersecurity defenses. Through the constant advancement of security protocols, ethical concerns, and responsible data handling, we may leverage the potential of GANs to improve cybersecurity and defend digital landscapes against new threats.

\bibliographystyle{IEEEtran}
\bibliography{cas-refs}

\begin{thebibliography}{100}
\providecommand{\url}[1]{#1}
\csname url@samestyle\endcsname
\providecommand{\newblock}{\relax}
\providecommand{\bibinfo}[2]{#2}
\providecommand{\BIBentrySTDinterwordspacing}{\spaceskip=0pt\relax}
\providecommand{\BIBentryALTinterwordstretchfactor}{4}
\providecommand{\BIBentryALTinterwordspacing}{\spaceskip=\fontdimen2\font plus
\BIBentryALTinterwordstretchfactor\fontdimen3\font minus \fontdimen4\font\relax}
\providecommand{\BIBforeignlanguage}[2]{{%
\expandafter\ifx\csname l@#1\endcsname\relax
\typeout{** WARNING: IEEEtran.bst: No hyphenation pattern has been}%
\typeout{** loaded for the language `#1'. Using the pattern for}%
\typeout{** the default language instead.}%
\else
\language=\csname l@#1\endcsname
\fi
#2}}
\providecommand{\BIBdecl}{\relax}
\BIBdecl

\bibitem{mao2020perigee}
Y.~Mao, S.~Deb, S.~B. Venkatakrishnan, S.~Kannan, and K.~Srinivasan, ``Perigee: Efficient peer-to-peer network design for blockchains,'' in \emph{Proceedings of the 39th Symposium on Principles of Distributed Computing}, 2020, pp. 428--437.

\bibitem{xue2023goldfish}
B.~Xue, Y.~Mao, S.~B. Venkatakrishnan, and S.~Kannan, ``Goldfish: Peer selection using matrix completion in unstructured p2p network,'' in \emph{2023 IEEE International Conference on Blockchain and Cryptocurrency (ICBC)}.\hskip 1em plus 0.5em minus 0.4em\relax IEEE, 2023, pp. 1--9.

\bibitem{baur2019cyber}
S.~Baur-Yazbeck, J.~Frickenstein, and D.~Medine, ``Cyber security in financial sector development,'' \emph{CGAP Background Documents}, vol.~5, no.~2, 2019.

\bibitem{dou2023towards}
F.~Dou, J.~Ye, G.~Yuan, Q.~Lu, W.~Niu, H.~Sun, L.~Guan, G.~Lu, G.~Mai, N.~Liu \emph{et~al.}, ``Towards artificial general intelligence (agi) in the internet of things (iot): Opportunities and challenges,'' \emph{arXiv preprint arXiv:2309.07438}, 2023.

\bibitem{mao2023less}
Y.~Mao and S.~B. Venkatakrishnan, ``Less is more: Understanding network bias in proof-of-work blockchains,'' \emph{Mathematics}, vol.~11, no.~23, p. 4741, 2023.

\bibitem{daricili2022national}
A.~B. Daricili and S.~Celik, ``National security 2.0: The cyber security of critical infrastructure,'' \emph{PERCEPTIONS: Journal of International Affairs}, vol.~26, no.~2, pp. 259--276, 2022.

\bibitem{carr2019into}
T.~Carr, J.~Zhuang, D.~Sablan, E.~LaRue, Y.~Wu, M.~Al~Hasan, and G.~Mohler, ``Into the reverie: Exploration of the dream market,'' in \emph{2019 IEEE International Conference on Big Data (Big Data)}.\hskip 1em plus 0.5em minus 0.4em\relax IEEE, 2019, pp. 1432--1441.

\bibitem{ali2021looking}
R.~Ali, ``Looking to the future of the cyber security landscape,'' \emph{Network Security}, vol. 2021, no.~3, pp. 8--10, 2021.

\bibitem{zhuang2022geometrically}
J.~Zhuang and D.~Wang, ``Geometrically matched multi-source microscopic image synthesis using bidirectional adversarial networks,'' in \emph{Proceedings of 2021 International Conference on Medical Imaging and Computer-Aided Diagnosis (MICAD 2021) Medical Imaging and Computer-Aided Diagnosis}.\hskip 1em plus 0.5em minus 0.4em\relax Springer, 2022, pp. 79--88.

\bibitem{mo2022trafficflowgan}
Z.~Mo, Y.~Fu, D.~Xu, and X.~Di, ``Trafficflowgan: Physics-informed flow based generative adversarial network for uncertainty quantification,'' in \emph{Joint European Conference on Machine Learning and Knowledge Discovery in Databases}.\hskip 1em plus 0.5em minus 0.4em\relax Springer, 2022, pp. 323--339.

\bibitem{zhuang2022defending}
J.~Zhuang and M.~Al~Hasan, ``Defending graph convolutional networks against dynamic graph perturbations via bayesian self-supervision,'' in \emph{Proceedings of the AAAI Conference on Artificial Intelligence}, vol.~36, no.~4, 2022, pp. 4405--4413.

\bibitem{zhuang2022robust}
------, ``Robust node classification on graphs: Jointly from bayesian label transition and topology-based label propagation,'' in \emph{Proceedings of the 31st ACM International Conference on Information \& Knowledge Management}, 2022, pp. 2795--2805.

\bibitem{mazaed2022multifaceted}
F.~Mazaed~Alotaibi and Fawad, ``A multifaceted deep generative adversarial networks model for mobile malware detection,'' \emph{Applied Sciences}, vol.~12, no.~19, p. 9403, 2022.

\bibitem{chen2023explainable}
J.~Chen, L.~Zhang, J.~Riem, G.~Adam, N.~D. Bastian, and T.~Lan, ``Explainable learning-based intrusion detection supported by memristors,'' in \emph{2023 IEEE Conference on Artificial Intelligence (CAI)}.\hskip 1em plus 0.5em minus 0.4em\relax IEEE, 2023, pp. 195--196.

\bibitem{ruan2022causal}
K.~Ruan, J.~Zhang, X.~Di, and E.~Bareinboim, ``Causal imitation learning via inverse reinforcement learning,'' in \emph{The Eleventh International Conference on Learning Representations}, 2022.

\bibitem{zhou2022adversarial}
S.~Zhou, C.~Liu, D.~Ye, T.~Zhu, W.~Zhou, and P.~S. Yu, ``Adversarial attacks and defenses in deep learning: From a perspective of cybersecurity,'' \emph{ACM Computing Surveys}, vol.~55, no.~8, pp. 1--39, 2022.

\bibitem{lyu2024task}
W.~Lyu, X.~Lin, S.~Zheng, L.~Pang, H.~Ling, S.~Jha, and C.~Chen, ``Task-agnostic detector for insertion-based backdoor attacks,'' \emph{arXiv preprint arXiv:2403.17155}, 2024.

\bibitem{zhuang2022does}
J.~Zhuang and M.~A. Hasan, ``How does bayesian noisy self-supervision defend graph convolutional networks?'' \emph{Neural Processing Letters}, vol.~54, no.~4, pp. 2997--3018, 2022.

\bibitem{lyu2023backdoor}
W.~Lyu, S.~Zheng, H.~Ling, and C.~Chen, ``Backdoor attacks against transformers with attention enhancement,'' in \emph{ICLR 2023 Workshop on Backdoor Attacks and Defenses in Machine Learning}, 2023.

\bibitem{zhuang2022deperturbation}
J.~Zhuang and M.~Al~Hasan, ``Deperturbation of online social networks via bayesian label transition,'' in \emph{Proceedings of the 2022 SIAM International Conference on Data Mining (SDM)}.\hskip 1em plus 0.5em minus 0.4em\relax SIAM, 2022, pp. 603--611.

\bibitem{singh2021survey}
J.~Singh and J.~Singh, ``A survey on machine learning-based malware detection in executable files,'' \emph{Journal of Systems Architecture}, 2021.

\bibitem{liu2019machine}
H.~Liu and B.~Lang, ``Machine learning and deep learning methods for intrusion detection systems: A survey,'' \emph{applied sciences}, vol.~9, no.~20, p. 4396, 2019.

\bibitem{zhuang2021non}
J.~Zhuang and M.~Al~Hasan, ``Non-exhaustive learning using gaussian mixture generative adversarial networks,'' in \emph{Machine Learning and Knowledge Discovery in Databases. Research Track: European Conference, ECML PKDD 2021, Bilbao, Spain, September 13--17, 2021, Proceedings, Part II 21}.\hskip 1em plus 0.5em minus 0.4em\relax Springer, 2021, pp. 3--18.

\bibitem{chen2023ride}
J.~Chen, L.~Zhang, J.~Riem, G.~Adam, N.~D. Bastian, and T.~Lan, ``Ride: Real-time intrusion detection via explainable machine learning implemented in a memristor hardware architecture,'' in \emph{2023 IEEE Conference on Dependable and Secure Computing (DSC)}.\hskip 1em plus 0.5em minus 0.4em\relax IEEE, 2023, pp. 1--8.

\bibitem{zeng2019deep}
Y.~Zeng, H.~Gu, W.~Wei, and Y.~Guo, ``$ deep-full-range $: a deep learning based network encrypted traffic classification and intrusion detection framework,'' \emph{IEEE Access}, vol.~7, pp. 45\,182--45\,190, 2019.

\bibitem{mo2022uncertainty}
Z.~Mo and X.~Di, ``Uncertainty quantification of car-following behaviors: physics-informed generative adversarial networks,'' in \emph{the 28th ACM SIGKDD in conjunction with the 11th International Workshop on Urban Computing (UrbComp2022)}, 2022.

\bibitem{goodfellow2014generative}
I.~Goodfellow, J.~Pouget-Abadie, M.~Mirza, B.~Xu, D.~Warde-Farley, S.~Ozair, A.~Courville, and Y.~Bengio, ``Generative adversarial nets,'' \emph{Advances in neural information processing systems}, vol.~27, 2014.

\bibitem{feng2018semi}
Z.~Feng, D.~Nie, L.~Wang, and D.~Shen, ``Semi-supervised learning for pelvic mr image segmentation based on multi-task residual fully convolutional networks,'' in \emph{2018 IEEE 15th International Symposium on Biomedical Imaging (ISBI 2018)}.\hskip 1em plus 0.5em minus 0.4em\relax IEEE, 2018.

\bibitem{feng2021two}
Z.~Feng, J.~A. Sivak, and A.~K. Krishnamurthy, ``Two-stream attention spatio-temporal network for classification of echocardiography videos,'' in \emph{2021 IEEE 18th International Symposium on Biomedical Imaging (ISBI)}.\hskip 1em plus 0.5em minus 0.4em\relax IEEE, 2021, pp. 1461--1465.

\bibitem{zhang2024deepgi}
Y.~Zhang, Y.~Gong, D.~Cui, X.~Li, and X.~Shen, ``Deepgi: An automated approach for gastrointestinal tract segmentation in mri scans,'' \emph{arXiv preprint arXiv:2401.15354}, 2024.

\bibitem{creswell2018generative}
A.~Creswell, T.~White, V.~Dumoulin, K.~Arulkumaran, B.~Sengupta, and A.~A. Bharath, ``Generative adversarial networks: An overview,'' \emph{IEEE signal processing magazine}, vol.~35, no.~1, pp. 53--65, 2018.

\bibitem{durgadevi2021generative}
M.~Durgadevi \emph{et~al.}, ``Generative adversarial network (gan): a general review on different variants of gan and applications,'' in \emph{2021 6th International Conference on Communication and Electronics Systems (ICCES)}.\hskip 1em plus 0.5em minus 0.4em\relax IEEE, 2021, pp. 1--8.

\bibitem{mo2022quantifying}
Z.~Mo, Y.~Fu, and X.~Di, ``Quantifying uncertainty in traffic state estimation using generative adversarial networks,'' in \emph{2022 IEEE 25th International Conference on Intelligent Transportation Systems (ITSC)}.\hskip 1em plus 0.5em minus 0.4em\relax IEEE, 2022, pp. 2769--2774.

\bibitem{yan2019method}
X.~Yan, B.~Cui, Y.~Xu, P.~Shi, and Z.~Wang, ``A method of information protection for collaborative deep learning under gan model attack,'' \emph{IEEE/ACM Transactions on Computational Biology and Bioinformatics}, vol.~18, no.~3, pp. 871--881, 2019.

\bibitem{hu2022generating}
W.~Hu and Y.~Tan, ``Generating adversarial malware examples for black-box attacks based on gan,'' in \emph{International Conference on Data Mining and Big Data}.\hskip 1em plus 0.5em minus 0.4em\relax Springer, 2022, pp. 409--423.

\bibitem{t15_arora2022review}
A.~Arora and Shantanu, ``A review on application of gans in cybersecurity domain,'' \emph{IETE Technical Review}, 2022.

\bibitem{chen2023real}
J.~Chen, H.~Zhou, Y.~Mei, G.~Adam, N.~D. Bastian, and T.~Lan, ``Real-time network intrusion detection via decision transformers,'' \emph{arXiv preprint arXiv:2312.07696}, 2023.

\bibitem{AS2_dunmore2023comprehensive}
A.~Dunmore, J.~Jang-Jaccard, F.~Sabrina, and J.~Kwak, ``A comprehensive survey of generative adversarial networks (gans) in cybersecurity intrusion detection,'' \emph{IEEE Access}, 2023.

\bibitem{U9_moti2021generative}
Z.~Moti, S.~Hashemi, H.~Karimipour, A.~Dehghantanha, A.~N. Jahromi, L.~Abdi, and F.~Alavi, ``Generative adversarial network to detect unseen internet of things malware,'' \emph{Ad Hoc Networks}, vol. 122, p. 102591, 2021.

\bibitem{li2024comprehensive}
Z.~Li, H.~Zhu, H.~Liu, J.~Song, and Q.~Cheng, ``Comprehensive evaluation of mal-api-2019 dataset by machine learning in malware detection,'' \emph{arXiv preprint arXiv:2403.02232}, 2024.

\bibitem{U1_ling2023adversarial}
X.~Ling, L.~Wu, J.~Zhang, Z.~Qu, W.~Deng, X.~Chen, Y.~Qian, C.~Wu, S.~Ji, T.~Luo \emph{et~al.}, ``Adversarial attacks against windows pe malware detection: A survey of the state-of-the-art,'' \emph{Computers \& Security}, p. 103134, 2023.

\bibitem{ma2023learning}
H.~Ma, D.~Zeng, and Y.~Liu, ``Learning optimal group-structured individualized treatment rules with many treatments,'' \emph{Journal of Machine Learning Research}, vol.~24, no. 102, pp. 1--48, 2023.

\bibitem{ma2022learning}
------, ``Learning individualized treatment rules with many treatments: A supervised clustering approach using adaptive fusion,'' \emph{Advances in Neural Information Processing Systems}, vol.~35, pp. 15\,956--15\,969, 2022.

\bibitem{U4_xia2022gan}
X.~Xia, X.~Pan, N.~Li, X.~He, L.~Ma, X.~Zhang, and N.~Ding, ``Gan-based anomaly detection: A review,'' \emph{Neurocomputing}, vol. 493, pp. 497--535, 2022.

\bibitem{AS5_berman2019survey}
D.~S. Berman, A.~L. Buczak, J.~S. Chavis, and C.~L. Corbett, ``A survey of deep learning methods for cyber security,'' \emph{Information}, vol.~10, no.~4, p. 122, 2019.

\bibitem{wang2024balanced}
Y.~Wang, J.~Wu, N.~Hovakimyan, and R.~Sun, ``Balanced training for sparse gans,'' \emph{Advances in Neural Information Processing Systems}, vol.~36, 2024.

\bibitem{AS6_wu2020network}
Y.~Wu, D.~Wei, and J.~Feng, ``Network attacks detection methods based on deep learning techniques: a survey,'' \emph{Security and Communication Networks}, vol. 2020, pp. 1--17, 2020.

\bibitem{de2023artificial}
A.~J.~G. de~Azambuja, C.~Plesker, K.~Sch{\"u}tzer, R.~Anderl, B.~Schleich, and V.~R. Almeida, ``Artificial intelligence-based cyber security in the context of industry 4.0—a survey,'' \emph{Electronics}, vol.~12, no.~8, p. 1920, 2023.

\bibitem{balaji2019conditional}
Y.~Balaji, M.~R. Min, B.~Bai, R.~Chellappa, and H.~P. Graf, ``Conditional gan with discriminative filter generation for text-to-video synthesis.'' in \emph{IJCAI}, vol.~1, no. 2019, 2019, p.~2.

\bibitem{AS1_randhawa2021security}
R.~H. Randhawa, N.~Aslam, M.~Alauthman, H.~Rafiq, and F.~Comeau, ``Security hardening of botnet detectors using generative adversarial networks,'' \emph{IEEE Access}, vol.~9, pp. 78\,276--78\,292, 2021.

\bibitem{t1_duy2021digfupas}
P.~T. Duy, N.~H. Khoa, A.~G.-T. Nguyen, V.-H. Pham \emph{et~al.}, ``Digfupas: Deceive ids with gan and function-preserving on adversarial samples in sdn-enabled networks,'' \emph{Computers \& Security}, vol. 109, p. 102367, 2021.

\bibitem{AS3_zhang2020brute}
S.~Zhang, X.~Xie, and Y.~Xu, ``A brute-force black-box method to attack machine learning-based systems in cybersecurity,'' \emph{IEEE Access}, vol.~8, pp. 128\,250--128\,263, 2020.

\bibitem{AS7_kim2022obfuscated}
J.-Y. Kim and S.-B. Cho, ``Obfuscated malware detection using deep generative model based on global/local features,'' \emph{Computers \& Security}, vol. 112, p. 102501, 2022.

\bibitem{AS4_chui2023three}
K.~T. Chui, B.~B. Gupta, P.~Chaurasia, V.~Arya, A.~Almomani, and W.~Alhalabi, ``Three-stage data generation algorithm for multiclass network intrusion detection with highly imbalanced dataset,'' \emph{International Journal of Intelligent Networks}, vol.~4, pp. 202--210, 2023.

\bibitem{AS11_kumar2023synthetic}
V.~Kumar and D.~Sinha, ``Synthetic attack data generation model applying generative adversarial network for intrusion detection,'' \emph{Computers \& Security}, vol. 125, p. 103054, 2023.

\bibitem{AS8_won2022plausmal}
D.-O. Won, Y.-N. Jang, and S.-W. Lee, ``Plausmal-gan: Plausible malware training based on generative adversarial networks for analogous zero-day malware detection,'' \emph{IEEE Transactions on Emerging Topics in Computing}, vol.~11, no.~1, pp. 82--94, 2022.

\bibitem{AS9_reilly2023robustness}
C.~Reilly, S.~O~Shaughnessy, and C.~Thorpe, ``Robustness of image-based malware classification models trained with generative adversarial networks,'' in \emph{Proceedings of the 2023 European Interdisciplinary Cybersecurity Conference}, 2023, pp. 92--99.

\bibitem{U8_lu2019generative}
Y.~Lu and J.~Li, ``Generative adversarial network for improving deep learning based malware classification,'' in \emph{2019 Winter Simulation Conference (WSC)}.\hskip 1em plus 0.5em minus 0.4em\relax IEEE, 2019, pp. 584--593.

\bibitem{de2020intrusion}
P.~F. de~Araujo-Filho, G.~Kaddoum, D.~R. Campelo, A.~G. Santos, D.~Mac{\^e}do, and C.~Zanchettin, ``Intrusion detection for cyber--physical systems using generative adversarial networks in fog environment,'' \emph{IEEE Internet of Things Journal}, vol.~8, no.~8, pp. 6247--6256, 2020.

\bibitem{t6_shahriar2020g}
M.~H. Shahriar, N.~I. Haque, M.~A. Rahman, and M.~Alonso, ``G-ids: Generative adversarial networks assisted intrusion detection system,'' in \emph{2020 IEEE 44th Annual Computers, Software, and Applications Conference (COMPSAC)}.\hskip 1em plus 0.5em minus 0.4em\relax IEEE, 2020, pp. 376--385.

\bibitem{chen2016infogan}
X.~Chen, Y.~Duan, R.~Houthooft, J.~Schulman, I.~Sutskever, and P.~Abbeel, ``Infogan: Interpretable representation learning by information maximizing generative adversarial nets,'' \emph{Advances in neural information processing systems}, vol.~29, 2016.

\bibitem{t9_ferrag2021federated}
M.~A. Ferrag, O.~Friha, L.~Maglaras, H.~Janicke, and L.~Shu, ``Federated deep learning for cyber security in the internet of things: Concepts, applications, and experimental analysis,'' \emph{IEEE Access}, vol.~9, pp. 138\,509--138\,542, 2021.

\bibitem{t11_das2022fgan}
S.~Das, ``Fgan: Federated generative adversarial networks for anomaly detection in network traffic,'' \emph{arXiv preprint arXiv:2203.11106}.

\bibitem{yang2022network}
Y.~Yang, C.~Yao, J.~Yang, and K.~Yin, ``A network security situation element extraction method based on conditional generative adversarial network and transformer,'' \emph{IEEE Access}, vol.~10, pp. 107\,416--107\,430, 2022.

\bibitem{t4_schneider2023dual}
J.~Schneider and G.~Apruzzese, ``Dual adversarial attacks: Fooling humans and classifiers,'' \emph{Journal of Information Security and Applications}, vol.~75, p. 103502, 2023.

\bibitem{zhuang2023robust}
J.~Zhuang and M.~A. Hasan, ``Robust node representation learning via graph variational diffusion networks,'' \emph{arXiv preprint arXiv:2312.10903}, 2023.

\bibitem{t12_lucas2023adversarial}
K.~Lucas, S.~Pai, W.~Lin, L.~Bauer, M.~K. Reiter, and M.~Sharif, ``Adversarial training for $\{$Raw-Binary$\}$ malware classifiers,'' in \emph{32nd USENIX Security Symposium (USENIX Security 23)}, 2023.

\bibitem{laykaviriyakul2023collaborative}
P.~Laykaviriyakul and E.~Phaisangittisagul, ``Collaborative defense-gan for protecting adversarial attacks on classification system,'' \emph{Expert Systems with Applications}, vol. 214, p. 118957, 2023.

\bibitem{t8_ge2023explainable}
W.~Ge, J.~Wang, T.~Lin, B.~Tang, and X.~Li, ``Explainable cyber threat behavior identification based on self-adversarial topic generation,'' \emph{Computers \& Security}, vol. 132, p. 103369, 2023.

\bibitem{t20_lee2021gan}
J.~Lee and K.~Park, ``Gan-based imbalanced data intrusion detection system,'' \emph{Personal and Ubiquitous Computing}, 2021.

\bibitem{zhang2020multiscale}
Q.~Zhang, C.~D. Heldermon, and C.~Toler-Franklin, ``Multiscale detection of cancerous tissue in high resolution slide scans,'' in \emph{International Symposium on Visual Computing}.\hskip 1em plus 0.5em minus 0.4em\relax Springer, 2020, pp. 139--153.

\bibitem{bian2022learnable}
W.~Bian, Q.~Zhang, X.~Ye, and Y.~Chen, ``A learnable variational model for joint multimodal mri reconstruction and synthesis,'' in \emph{International Conference on Medical Image Computing and Computer-Assisted Intervention}.\hskip 1em plus 0.5em minus 0.4em\relax Springer, 2022, pp. 354--364.

\bibitem{kar2022integrated}
J.~Kar and A.~Chakrabortty, ``An integrated generative adversarial network for identification and mitigation of cyber-attacks in wide-area control,'' in \emph{2022 IEEE Power \& Energy Society General Meeting (PESGM)}.\hskip 1em plus 0.5em minus 0.4em\relax IEEE, 2022, pp. 1--5.

\bibitem{t5_randhawa2022evasion}
R.~H. Randhawa, N.~Aslam, M.~Alauthman, and H.~Rafiq, ``Evasion generative adversarial network for low data regimes,'' \emph{IEEE Transactions on Artificial Intelligence}, 2022.

\bibitem{huang2022android}
Y.~Huang, X.~Li, M.~Qiao, K.~Tang, C.~Zhang, H.~Gui, P.~Wang, and F.~Liu, ``Android-sem: Generative adversarial network for android malware semantic enhancement model based on transfer learning,'' \emph{Electronics}, vol.~11, no.~5, p. 672, 2022.

\bibitem{t2_liu2020efficient}
Z.~Liu, S.~Li, Y.~Zhang, X.~Yun, and Z.~Cheng, ``Efficient malware originated traffic classification by using generative adversarial networks,'' in \emph{2020 IEEE symposium on computers and communications (ISCC)}.\hskip 1em plus 0.5em minus 0.4em\relax IEEE, 2020, pp. 1--7.

\bibitem{t14_wang2022evilmodel}
Z.~Wang, C.~Liu, X.~Cui, J.~Yin, and X.~Wang, ``Evilmodel 2.0: bringing neural network models into malware attacks,'' \emph{Computers \& Security}, vol. 120, p. 102807, 2022.

\bibitem{soleymanzadeh2022stable}
R.~Soleymanzadeh and R.~Kashef, ``A stable generative adversarial network architecture for network intrusion detection,'' in \emph{2022 IEEE International Conference on Cyber Security and Resilience (CSR)}.\hskip 1em plus 0.5em minus 0.4em\relax IEEE, 2022, pp. 9--15.

\bibitem{t7_park2022enhanced}
C.~Park, J.~Lee, Y.~Kim, J.-G. Park, H.~Kim, and D.~Hong, ``An enhanced ai-based network intrusion detection system using generative adversarial networks,'' \emph{IEEE Internet of Things Journal}, vol.~10, no.~3, pp. 2330--2345, 2022.

\bibitem{yin2018enhancing}
C.~Yin, Y.~Zhu, S.~Liu, J.~Fei, and H.~Zhang, ``An enhancing framework for botnet detection using generative adversarial networks,'' in \emph{2018 International Conference on Artificial Intelligence and Big Data (ICAIBD)}.\hskip 1em plus 0.5em minus 0.4em\relax IEEE, 2018, pp. 228--234.

\bibitem{t10_gupta2023chatgpt}
M.~Gupta, C.~Akiri, K.~Aryal, E.~Parker, and L.~Praharaj, ``From chatgpt to threatgpt: Impact of generative ai in cybersecurity and privacy,'' \emph{IEEE Access}, 2023.

\bibitem{hao2021asymmetric}
X.~Hao, W.~Ren, R.~Xiong, T.~Zhu, and K.-K.~R. Choo, ``Asymmetric cryptographic functions based on generative adversarial neural networks for internet of things,'' \emph{Future Generation Computer Systems}, vol. 124, pp. 243--253, 2021.

\bibitem{U25_zeng2020openattack}
G.~Zeng, F.~Qi, Q.~Zhou, T.~Zhang, Z.~Ma, B.~Hou, Y.~Zang, Z.~Liu, and M.~Sun, ``Openattack: An open-source textual adversarial attack toolkit,'' \emph{arXiv preprint arXiv:2009.09191}, 2020.

\bibitem{U26_papernot2016technical}
N.~Papernot, F.~Faghri, N.~Carlini, I.~Goodfellow, R.~Feinman, A.~Kurakin, C.~Xie, Y.~Sharma, T.~Brown, A.~Roy \emph{et~al.}, ``Technical report on the cleverhans v2. 1.0 adversarial examples library,'' \emph{arXiv preprint arXiv:1610.00768}, 2016.

\bibitem{U27_ling2019deepsec}
X.~Ling, S.~Ji, J.~Zou, J.~Wang, C.~Wu, B.~Li, and T.~Wang, ``Deepsec: A uniform platform for security analysis of deep learning model,'' in \emph{2019 IEEE symposium on security and privacy (SP)}.\hskip 1em plus 0.5em minus 0.4em\relax IEEE, 2019, pp. 673--690.

\bibitem{U28_li2020deeprobust}
Y.~Li, W.~Jin, H.~Xu, and J.~Tang, ``Deeprobust: A pytorch library for adversarial attacks and defenses,'' \emph{arXiv preprint arXiv:2005.06149}, 2020.

\bibitem{li2024mapping}
Z.~Li, B.~Guan, Y.~Wei, Y.~Zhou, J.~Zhang, and J.~Xu, ``Mapping new realities: Ground truth image creation with pix2pix image-to-image translation,'' \emph{arXiv preprint arXiv:2404.19265}, 2024.

\bibitem{babaagba2023evolutionary}
K.~O. Babaagba and J.~Wylie, ``An evolutionary based generative adversarial network inspired approach to defeating metamorphic malware,'' in \emph{Proceedings of the Companion Conference on Genetic and Evolutionary Computation}, 2023, pp. 1753--1759.

\bibitem{t3_ge2023explainable}
W.~Ge, J.~Wang, T.~Lin, B.~Tang, and X.~Li, ``Explainable cyber threat behavior identification based on self-adversarial topic generation,'' \emph{Computers \& Security}, vol. 132, p. 103369, 2023.

\bibitem{U7_chale2022generating}
M.~Chal{\'e} and N.~D. Bastian, ``Generating realistic cyber data for training and evaluating machine learning classifiers for network intrusion detection systems,'' \emph{Expert Systems with Applications}, vol. 207, p. 117936, 2022.

\bibitem{U5_rathore2023adversarial}
H.~Rathore, A.~Nandanwar, S.~K. Sahay, and M.~Sewak, ``Adversarial superiority in android malware detection: Lessons from reinforcement learning based evasion attacks and defenses,'' \emph{Forensic Science International: Digital Investigation}, vol.~44, p. 301511, 2023.

\bibitem{U12_bakhsh2023enhancing}
S.~A. Bakhsh, M.~A. Khan, F.~Ahmed, M.~S. Alshehri, H.~Ali, and J.~Ahmad, ``Enhancing iot network security through deep learning-powered intrusion detection system,'' \emph{Internet of Things}, vol.~24, p. 100936, 2023.

\bibitem{U10_dirgantoro2020generative}
K.~P. Dirgantoro, J.~M. Lee, and D.-S. Kim, ``Generative adversarial networks based on edge computing with blockchain architecture for security system,'' in \emph{2020 International Conference on Artificial Intelligence in Information and Communication (ICAIIC)}.\hskip 1em plus 0.5em minus 0.4em\relax IEEE, 2020, pp. 039--042.

\bibitem{feng2020performance}
R.~Feng, S.~Chen, X.~Xie, G.~Meng, S.-W. Lin, and Y.~Liu, ``A performance-sensitive malware detection system using deep learning on mobile devices,'' \emph{IEEE Transactions on Information Forensics and Security}, vol.~16, pp. 1563--1578, 2020.

\bibitem{amro2018malware}
B.~Amro, ``Malware detection techniques for mobile devices,'' \emph{arXiv preprint arXiv:1801.02837}, 2018.

\bibitem{shabtai2010malware}
A.~Shabtai, ``Malware detection on mobile devices,'' in \emph{2010 Eleventh International Conference on Mobile Data Management}.\hskip 1em plus 0.5em minus 0.4em\relax IEEE, 2010, pp. 289--290.

\bibitem{U6_xu2023gendroid}
G.~Xu, H.~Shao, J.~Cui, H.~Bai, J.~Li, G.~Bai, S.~Liu, W.~Meng, and X.~Zheng, ``Gendroid: A query-efficient black-box android adversarial attack framework,'' \emph{Computers \& Security}, p. 103359, 2023.

\bibitem{U11_ferrag2023generative}
M.~A. Ferrag, M.~Debbah, and M.~Al-Hawawreh, ``Generative ai for cyber threat-hunting in 6g-enabled iot networks,'' \emph{arXiv preprint arXiv:2303.11751}, 2023.

\bibitem{U13_de2020intrusion}
P.~F. de~Araujo-Filho, G.~Kaddoum, D.~R. Campelo, A.~G. Santos, D.~Mac{\^e}do, and C.~Zanchettin, ``Intrusion detection for cyber--physical systems using generative adversarial networks in fog environment,'' \emph{IEEE Internet of Things Journal}, vol.~8, no.~8, pp. 6247--6256, 2020.

\bibitem{zhuang2019lighter}
J.~Zhuang, M.~Gao, and M.~A. Hasan, ``Lighter u-net for segmenting white matter hyperintensities in mr images,'' in \emph{Proceedings of the 16th EAI International Conference on Mobile and Ubiquitous Systems: Computing, Networking and Services}, 2019, pp. 535--539.

\bibitem{t13_liu2023decompiling}
Z.~Liu, Y.~Yuan, S.~Wang, X.~Xie, and L.~Ma, ``Decompiling x86 deep neural network executables,'' in \emph{32nd USENIX Security Symposium (USENIX Security 23)}, 2023, pp. 7357--7374.

\bibitem{U14_nie2021intrusion}
L.~Nie, Y.~Wu, X.~Wang, L.~Guo, G.~Wang, X.~Gao, and S.~Li, ``Intrusion detection for secure social internet of things based on collaborative edge computing: a generative adversarial network-based approach,'' \emph{IEEE Transactions on Computational Social Systems}, vol.~9, no.~1, pp. 134--145, 2021.

\bibitem{zhang2019extra}
Q.~Zhang and Y.~Chen, ``Extra proximal-gradient inspired non-local network,'' \emph{arXiv preprint arXiv:1911.07144}, 2019.

\bibitem{zhang2020novel}
Q.~Zhang, X.~Ye, H.~Liu, and Y.~Chen, ``A novel learnable gradient descent type algorithm for non-convex non-smooth inverse problems,'' \emph{arXiv preprint arXiv:2003.06748}, 2020.

\bibitem{zhang2022extra}
Q.~Zhang, X.~Ye, and Y.~Chen, ``Extra proximal-gradient network with learned regularization for image compressive sensing reconstruction,'' \emph{Journal of Imaging}, vol.~8, no.~7, p. 178, 2022.

\bibitem{U3_zhu2023adfl}
C.~Zhu, J.~Zhang, X.~Sun, B.~Chen, and W.~Meng, ``Adfl: Defending backdoor attacks in federated learning via adversarial distillation,'' \emph{Computers \& Security}, p. 103366, 2023.

\bibitem{garcia2020iot23}
\BIBentryALTinterwordspacing
S.~Garcia, A.~Parmisano, and M.~J. Erquiaga, ``{IoT-23: A labeled dataset with malicious and benign IoT network traffic (Version 1.0.0)},'' Zenodo, 2020. [Online]. Available: \url{http://doi.org/10.5281/zenodo.4743746}
\BIBentrySTDinterwordspacing

\bibitem{t17_yilmaz2020addressing}
I.~Yilmaz, R.~Masum, and A.~Siraj, ``Addressing imbalanced data problem with generative adversarial network for intrusion detection,'' in \emph{2020 IEEE 21st international conference on information reuse and integration for data science (IRI)}.\hskip 1em plus 0.5em minus 0.4em\relax IEEE, 2020.

\bibitem{t18_liu2021gan}
X.~Liu, T.~Li, R.~Zhang, D.~Wu, Y.~Liu, and Z.~Yang, ``A gan and feature selection-based oversampling technique for intrusion detection,'' \emph{Security and Communication Networks}, 2021.

\bibitem{t19_nazari2021using}
E.~Nazari, P.~Branco, and G.-V. Jourdan, ``Using cgan to deal with class imbalance and small sample size in cybersecurity problems,'' in \emph{2021 18th International Conference on Privacy, Security and Trust (PST)}.\hskip 1em plus 0.5em minus 0.4em\relax IEEE, 2021, pp. 1--10.

\end{thebibliography}

\end{document}